\documentclass[a4paper,11pt]{article}
\usepackage[utf8]{inputenc}
\usepackage{color}
\usepackage{cite}
\usepackage{hyperref}
\usepackage{amsmath}
\usepackage{graphicx}
\renewcommand{\thesection}{\arabic{section}.}

\usepackage[top=0.4in, bottom=0.7in, left=0.60in, right=0.5in]{geometry}

\begin{document}
\title{{\bf\Large{Radiative decays of heavy-light quarkonia through $M1$, and $E1$ transitions in the framework of Bethe-Salpeter equation}}}
\author{Shashank Bhatnagar$^1$, Eshete Gebrehana$^2$}

\maketitle \small{$^1$Department of Physics, University Institute of Sciences, Chandigarh University, Mohali-140413, India\\}
$^2$Department of Physics, Addis Ababa University, P.O.Box 1176 Addis Ababa, Ethiopia\\

\begin{abstract}
\normalsize{In this work we study the radiative decays of heavy-light quarkonia through M1 and E1 transitions that involve quark-triangle diagrams with two hadron vertices, and are difficult to evaluate in BSE-CIA. We have expressed the transition amplitude $M_{fi}$ as a linear superposition of terms involving all possible combinations of $++$, and $--$ components of Salpeter wave functions of final and initial hadron, with coefficients being related to results of pole integrations over complex $\sigma$-plane.  We evaluate the decay widths for $M1$ transitions ($^3S_1 \rightarrow ^1S_0 +\gamma$), and $E1$ transitions ($^3S_1 \rightarrow ^1P_0 +\gamma$ and $^1P_0 \rightarrow ^3S_1 +\gamma$). We have used algebraic forms of Salpeter wave functions obtained through analytic solutions of mass spectral equations for ground and excited states of $0^{++},1^{--}$, and $0^{-+}$ heavy-light quarkonia in approximate harmonic oscillator basis to calculate their decay widths. The input parameters used by us were obtained by fitting to their mass spectra. We have compared our results with experimental data and other models, and found reasonable agreements.}
\end{abstract}

\section{Introduction}
The most important goals of hadronic physics is to bridge the gap between the QCD lagrangian, and the observed hadronic properties. One of the challenging areas in hadronic physics presently is probing the inner structure of hadrons. There has a been a renewed interest in recent
years in
spectroscopy of these heavy hadrons in charm and beauty sectors,
which was primarily due to experimental facilities the world over
such as BABAR, Belle, CLEO, DELPHI, BES etc.
\cite{Ecklund,babar09,belle10,cleo01,olive14}, which have been
providing accurate data on $c\overline{c}$, and $b\overline{b}$
hadrons with respect to their masses and decays. In the process
many new states have been discovered such as $\chi_{b0}(3P)$,
$\chi_{c0}(2P), X(3915), X(4260), X(4360), X(4430), X(4660)$
\cite{olive14}, some of which are exotic states, which can not be readily explained through the predictions of the quark model. The radiative
transitions of heavy quarkonia are of considerable experimental and theoretical interest, and provide an insight into the dynamics of quarkonium. The
radiative transitions between $0^{-+}$ (pseudoscalar), and $1^{--}$ (vector) mesons (for instance, $J/\Psi(nS) -> \eta_c(n'S) +\gamma$), which proceeds
through the emission of photon is characterized by $\Delta L=0$, there is change in C-parity between the initial and final hadron states, though the total C-parity is conserved. These are the magnetic dipole transition, $M1$. This transition mode is sensitive to relativistic effects, specially between different spatial multiplets ($n > n'$). The E1 transitions are characterized by $|\triangle L|=1$. Thus in these transitions, there is change in parity between the initial and final hadronic states, for instance, $\Psi(2S)\rightarrow \chi_c(1P) + \gamma$  or $\chi_c(1P) \rightarrow J/\Psi(1S) +\gamma$. In both M1 and E1 transitions, C-parity is conserved. Electric dipole transitions are much stronger than magnetic dipole transitions, and involve transitions between excited states. These transitions have been recently studied in various models, such as relativistic quark models \cite{eichten08, brambila04}, effective field theory\cite{brambila06,pineda13}, Light-front quark models \cite{vary18,Choi07}, Lattice QCD \cite{becirevic13,donald12}, Bethe-Salpeter equation \cite{mitra91,mitra01,karmanov15,hluf16}.

In this work we focus on the radiative decays of the charmed and bottom vector mesons through the processes, $V\rightarrow P\gamma$, $V\rightarrow S\gamma$, and $S\rightarrow V \gamma$, where, $V, P, S$ refer to vector, pseudoscalar and scalar quarkonia, and calculate the radiative decay widths of
$B^*$, and $D^*$ mesons for the above mentioned processes in the framework of $4\times 4$ Bethe-Salpeter equation. In our recent works
\cite{bhatnagar18, eshete19}, we had studied the mass spectrum of ground and excited states of heavy-light scalar ($0^{++}$), pseudoscalar ($0^{-+}$),
and vector ($1^{--}$) quarkonia, along with the leptonic decays of ground and excited states of $0^{-+}$, and $1^{--}$ quarkonia. These studies were
used to fit the input parameters of our model as $C_0$= 0.69, $\omega_0$= 0.22 GeV, $\Lambda_{QCD}$= 0.250 GeV, and $A_0$= 0.01, with input quark masses
$m_u$= 0.300 GeV, $m_s$= 0.430 GeV,$m_c$= 1.490 GeV, and $m_b$= 4.690 GeV. In the present work on radiative decays, we use these same input parameters to calculate the single photon decay widths for the above processes,

Now, as mentioned in our previous works \cite{bhatnagar18,eshete19,hluf16,hluf17}, we are not only interested in studying the mass
spectrum of hadrons, which no doubt is an important element to
study dynamics of hadrons, but also the hadronic wave functions
that play an important role in the calculation of decay constants,
form factors, structure functions etc. for $Q\overline{Q}$, and
$Q\overline{q}$ hadrons. These hadronic Bethe-Salpeter wave
functions were calculated algebraically by us in \cite{hluf16,bhatnagar18, eshete19}. The plots of these wave functions \cite{eshete19} show that they
can provide information not only about the long distance non-perturbative physics, but also act as a bridge between the long distance, and short
distance physics, and are provide us information about the contribution of the short ranged coulomb interactions in the mass spectral calculation of
heavy-light quarkonia. These wave functions and can also lead to studies on a number of processes involving
$Q\overline{Q}$, and $Q\overline{q}$ states, and provide a guide for future experiments.

This paper is organized as follows: In section 2, we introduce the
formulation of the $4\times 4$ Bethe-Salpeter equation under the
covariant instantaneous ansatz, and derive the hadron-quark
vertex. In sections 3, 4, and 5, we  calculate the single photon decay widths for the processes, $V\rightarrow P\gamma$, $V\rightarrow S\gamma$, and
$S\rightarrow V\gamma$, where, P, S, and V are the pseudoscalar, scalar and vector heavy-light quarkonium states. In section 6, we provide the numerical
results and discussion.

\section{Formulation of the BSE under CIA}
Our work is based on QCD motivated BSE in ladder approximation, which is an approximate description, with an effective four-fermion interaction mediated
by a gluonic propagator that serves as the kernel of BSE in the lowest order. The precise form of our kernel is taken in analogy with potential models,
which includes a confining term along with a one-gluon exchange term. Such effective forms of the BS kernel in ladder BSE have recently been used in
\cite{karmanov19,glwang,fredrico14,hea19,wang06}, and can predict bound states having a purely relativistic origin (as shown recently in
\cite{karmanov19}). As mentioned above, the BSE is quite general, and provides an effective description of bound quark- antiquark systems through a
suitable choice of input kernel for confinement.

The Bethe-Salpeter equation that describes the bound state of two quarks ($Q\overline{Q}$ or $Q\overline{q}$) of momenta $p_1$ and $p_2$, relative
momentum $q$, and meson momentum $P$ is
\begin{equation}
S_{F}^{-1}(p_{1})\Psi(P,q)S_{F}^{-1}(-p_{2}) =
i\int \frac{d^{4}q'}{(2\pi)^{4}}K(q,q')\Psi(P,q'),
\end{equation}
where $K(q,q')$ is the interaction kernel, and $S_{F}^{-1}(\pm p_{1,2})=\pm i{\not}p_{1,2}+ m_{1,2}$ are the usual quark and antiquark propagators.
We now make use of the Covariant Instantaneous Ansatz, where,
$K(q,q')=K(\widehat{q},\widehat{q}')$ on the BS kernel, where
$\widehat{q}_\mu= q_\mu- \frac{q.P}{P^2}P_\mu$ is the component of
internal momentum of the hadron that is orthogonal to the total
hadron momentum, i.e. $\widehat{q}.P=0$, while $\sigma
P_\mu=\frac{q.P}{P^2}P_\mu$ is the component of $q$ longitudinal
to $P$, where the 4-dimensional volume element is,
$d^4q=d^3\widehat{q}Md\sigma$, and following a sequence of steps
outlined in \cite{hluf16}, we get  four
Salpeter equations (in 4D variable $\widehat{q}$), which are
effective 3D forms of BSE (Salpeter equations) given below:

\begin{eqnarray}
 &&\nonumber(M-\omega_1-\omega_2)\psi^{++}(\hat{q})=\Lambda_{1}^{+}(\hat{q})\Gamma(\hat{q})\Lambda_{2}^{+}(\hat{q})\\&&
   \nonumber(M+\omega_1+\omega_2)\psi^{--}(\hat{q})=-\Lambda_{1}^{-}(\hat{q})\Gamma(\hat{q})\Lambda_{2}^{-}(\hat{q})\\&&
\nonumber \psi^{+-}(\hat{q})=0.\\&&
 \psi^{-+}(\hat{q})=0\label{fw5}
\end{eqnarray}

Thus, in our framework, an important role is played by the component, $\hat{q}_{\mu}$, which is always orthogonal to $P_{\mu}$  (i.e.$ \hat{q}.P=0$) regardless of whether $q_{\mu}$ is on-shell $(q.P=0)$, or off-shell $(q.P \neq 0)$.  Thus, in view of this remarkable property of $\hat{q}_{\mu}$, which makes it an effectively 3D vector, our twin objective of (i) 3D structure of BSE (Salpeter equations) as the controlling equation for spectra, and (ii) a general enough (off-shell) structure of BS vertex function, $\Gamma(\hat{q})$ to facilitate applications to transition amplitudes in 4D form is largely met if the BS kernel depends on $\hat{q}_{\mu}$. Thus, the ansatz, $K(q,q')=K(\widehat{q},\widehat{q}')$ on the BS kernel, is known as the Covariant Instantaneous Ansatz, and is a Lorentz-invariant genaralization of the Instantaneous Approximation (IA).

Following a sequence of steps, the 4D B.S. wave function can be written as \cite{hluf16}
\begin{equation}
 \Psi(P,\hat q)=S_1(p_1)\Gamma(\hat q)S_2(-p_2),
\end{equation}
where the 4D hadron-quark vertex, that enters into the definition of the 4D BS wave function in the previous equation, can be identified as,
\begin{equation}\label{6a}
 \Gamma(\hat q)=\int\frac{d^3\hat q'}{(2\pi)^3}K(\hat q,\hat q')\psi(\hat q')
\end{equation}

Here, we further wish to point out that the effectively 3D Salpeter equations (used for mass spectral calculations), as well as the vertex function, $\Gamma(\hat{q})$ in Eq.(2) (used for transition amplitude calculations) have a common dependence on the quantity, $\hat{q}^2$, whose most important property is its positive definiteness, $\hat{q}^2=q^2-\frac{(q.P)^2}{P^2}\geq 0$ on the hadron mass shell $(P^2=-M^2)$ throughout the entire 4D space\cite{mitra91,mitra01,wang06,bhatnagar14}. Thus, $|\hat{q}|=\sqrt{q^2-\frac{(q.P)^2}{P^2}}$ has been regarded as a Lorentz-invariant variable..

Thus all equations and wave functions have an explicit dependence on the 4D variable, $\hat{q}$. These equations form a zero-order basis for making contact with the mass spectrum of hadronic states, and calculations of various types of transition amplitudes through appropriate quark-loop diagrams. This is what makes these equations and wave functions 4D, but expressed in an effective 3D form. This increases the applicability of this framework of Covariant Instantaneous Ansatz all the way from low energy spectra to high energy transition amplitudes. Further, in this approach, the most important aspect is the appearance of hadron-quark vertex, $\Gamma(\hat{q})$ (used for calculation of transition amplitudes) on the right side of effectively 3D Salpeter equations (used for calculation of spectra) in Eq.(2), which  gives a dynamical link between low energy mass spectroscopy and high energy transition amplitudes. Such dynamical links between 3D spectra and 4D transition amplitudes have been explored in details in \cite{mitra91,mitra01,bhatnagar14}, by showing the exact interconnection between the 3D and 4D BSE.

The 4D B.S. wave function can be expressed in terms of the projected wave functions as
\begin{equation}\label{bb1}
\psi(\hat q)=\psi^{++}(\hat q)+\psi^{+-}(\hat q)+\psi^{-+}(\hat q)+\psi^{--}(\hat q),
\end{equation}
where
\begin{equation}\label{a7}
 \psi^{\pm\pm}(\hat q)= \Lambda^\pm_{1}(\hat q)\frac{{\not}P}{M}\psi(\hat q)
 \frac{{\not}P}{M}\Lambda^\pm_{2}(\hat q)
\end{equation}
and the projection operators
\begin{equation}\label{a1}
 \Lambda^\pm_{j}(\hat q)=\frac{1}{2\omega_j}\bigg[\frac{{\not}P}{M}\omega_j\pm J(j)(im_j+{\not}\hat q)\bigg],~~~J(j)=(-1)^{j+1},~~j=1,2
\end{equation}
with the relation
\begin{equation}
 \omega^2_j=m_j^2+\hat q^2
\end{equation}

\section{Radiative decays of heavy-light quarkonia through $V\rightarrow P\gamma$}
\begin{figure}[h!]
 \centering
 \includegraphics[width=12cm,height=6cm]{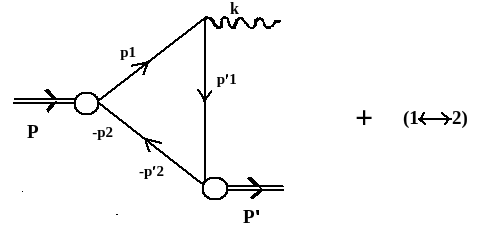}
 \caption{Radiative decays of heavy-light quarkonia}
 \label{fig:1}
\end{figure}

The single photon decay of vector ($1^{--}$) quarkonia is described by the direct and exchange Feynman diagrams as in Figure
\ref{fig:1}.

To apply the framework of BSE to study radiative decays, $V->P\gamma$, we have to remember that there are two Lorents frames, one the rest frame of the
initial meson, and the other, the rest frame of final meson. To calculate further, we first write relationship between the momentum variables of the initial and final meson. Here, $P$, and $q$ are the total momentum and the internal momentum of initial hadron, while $P'$, and $q'$ are the corresponding variables of the final hadron, and let $k$, and $\epsilon^{\lambda'}$ be momentum and polarization vectors of emitted photon, while $\epsilon^{\lambda}$ be the polarization vector of initial meson. Thus if $p_{1,2}$, and $p'_{1,2}$ are the momenta of the two quarks in initial and final hadron respectively, then,we have, the momentum relations:

\begin{eqnarray}
&&\nonumber P=p_1+p_2;  p_{1,2}=\hat{m}_{1,2}P\pm q\\&&
 P'=p'_1+p'_2;  p'_{1,2}=\hat{m}_{1,2}P'\pm q'
\end{eqnarray}

for initial and final hadrons respectively. From the Feynman diagrams we see that conservation of momentum demands that, $P= P'+k$, while from the first diagram, $p_1=p'_1+k$, and $-p_2=-p'_2$, where $k=P-P'$ is the momentum of the emitted photon. Making use of the above equations, we can express, the relationship between the internal momenta of the two hadrons in terms of the photon momentum, $k$ as,

\begin{equation}\label{9}
q'=q+(\hat{m}_1-1)k = q- \hat{m}_2 k,
\end{equation}

with $\hat{m}_{1,2}=\frac{1}{2}[1\pm\frac{(m^{2}_{1}-m^{2}_{2})}{M^{2}}]$ being the Wightman-Garding definitions\cite{bhatnagar14} of masses of individual quarks, which ensure that $P.q=0$ on the mass shells of either quarks, even when $m_1 \neq m_2$. They act like momentum partitioning functions for the two quarks in a hadron. We had already decomposed the internal momentum $q$ of the initial hadron into two components, $q=(\hat{q},iM\sigma)$, where $\hat{q}_{\mu}$ is the component of internal momentum transverse to $P$ such that $\hat{q}.P=0$, while $\sigma$ is the longitudinal component in the direction of $P$. Similarly for final meson, we decompose its internal momentum, $q'$ into two components $q'=(\hat{q}',iM\sigma')$, with $\hat{q}'=q'-\sigma'P$ transverse to initial hadron momentum, $P$, and $\sigma'=\frac{q'.P}{P^2}$, longitudinal to $P$. Thus, $P.\hat{q}'=0$. We now first try to find the relationship between the transverse components of internal momenta of the two hadrons, $\hat{q}$, and $\hat{q}'$. For this, we resolve all momenta in Eq.(\ref{9}) along the direction transverse to the momentum of the initial meson, $P$. Thus we can express Eq.(\ref{9}) as

\begin{eqnarray}
&&\nonumber \hat{q}'=\hat{q}+\hat{m}_2 (\hat{P}'-\hat{P}),\\&&
\nonumber \hat{P}=0\\&&
\hat{P}'=P'-\frac{P'.P}{P^2}P,
\end{eqnarray}

where, it can be easily checked that $\hat{P}.P=0$, and thus $\hat{P}'$ is orthogonal to $P$. The above equation can be simplified as,

\begin{equation}
\hat{q}'=\hat{q}+\hat{m}_2 (\hat{P}').
\end{equation}

It is to be mentioned that the above relation connecting $\hat{q}$, and $\hat{q}'$ is again consistent with the transversality of $\hat{q}'$ with $P$, given by the definition of $\hat{q}'$. And it can be easily checked that, $\hat{q}'.P=0$.

Now, the kinematics gets simplified in the rest frame of the initial meson, where we have $P=(\overrightarrow{0}, iM)$, while for emitted meson, $P'=(\overrightarrow{P}', iE')$, where $E'=\sqrt{\overrightarrow{P}'^2+M'^2}$, and since the photon momentum can be decomposed as, $k=(\overrightarrow{k},i|\overrightarrow{k}|)$, where $\overrightarrow{k}=-\overrightarrow{P}'$, since final meson and photon would be emitted in opposite directions. Hence we get, $|\overrightarrow{P}'|=|\overrightarrow{k}|=\frac{M^2-M'^2}{2M}$. Thus the energy of the emitted meson can be expressed as, $E'=\frac{M^2+M'^2}{2M}$.

Further the dot products of momenta of the initial and the emitted meson can be expressed as,

\begin{equation}
P'.P=-ME'=-\frac{M^2+M'^2}{2}
\end{equation}

Thus, it can be seen that, $-E'$ acts as the projection of $P'$ along the direction of initial hadron momentum, $P$. Now, we try to find relationship between the time components, $\sigma$ and $\sigma'$ of the two hadrons. Taking dot product of Eq.(\ref{9}) with $P$, the momentum of the initial hadron, we obtain,
\begin{equation}
P.q'=P.q-\hat{m}_{2}P.k.
\end{equation}

Making use of the above decomposition of internal momenta, we obtain the relation between the longitudinal components of internal momenta of the two hadrons as,

\begin{eqnarray}\label{13}
&&\nonumber \sigma'= \sigma+\alpha; \\&&
\alpha=\hat{m}_{2}\frac{M'^2-M^2}{2M^2},
\end{eqnarray}

which is again a consequence of the transversality of $\hat{q}'$ with initial hadron momentum, $P$. Thus, up to Eq.(\ref{13}), the kinematics is the same for all the three processes ($V\rightarrow P\gamma$, $V\rightarrow S \gamma$, and $S \rightarrow V \gamma$)
studied in this work.

It is to be noted that 4D BS wave functions of vector meson involved in the process, that is measured in its own rest frame is,

\begin{equation}
\Psi_{V}(P,q)=S_{F}(p_{1})\Gamma_V(\hat{q})S_F(-p_{2}),
\end{equation}

Similar wave function, $\Psi_{P}(P',q')$ can be written for pseudoscalar meson in its own rest frame. However, for transition amplitude calculation, we are choosing to do calculation in the rest frame of the initial meson. Thus, we write, $\Psi_{P}(P',q')=S_{F}(p'_{1})\Gamma_V(\hat{q}')S_F(-p'_{2})$, with $\hat{q}'$ defined earlier as $\hat{q}'=q'-\frac{q'.P}{P^2}P$, that is transverse to initial hadron momentum, $P$.

The EM transition amplitude of the process is
\begin{equation}\label{15}
 M_{fi}=-i\int \frac{d^4q}{(2\pi)^4}Tr[e_q\overline{\Psi}_{P}(P', q'){\not}\epsilon^{\lambda'}\Psi_{V}(P,q)
S_F^{-1}(-p_2)+ e_{\overline{Q}}\overline{\Psi}_{P}(P', q')S_F^{-1}(p_1)\Psi_{V}(P, q){\not}\epsilon^{\lambda'}],
\end{equation}
where the $M_{fi}$ is written in the rest frame of the initial hadron, and the relationship between $q'$, and $q$ is given by Eq.(10). Here, the first term corresponds to the first diagram, where the photon is emitted from the quark ($q$), while the second term corresponds to the second
diagram where the photon is emitted from the antiquark ($\overline{Q}$) in vector meson.

In the above expression, $\Psi_{P}$ and $\Psi_{V}$ are the 4D BS wave functions of
pseudoscalar and vector quarkonia involved in the process, and are
expressed above, while  $e_{q}$ , and $e_{Q}$ are the
electric charge of quark, and antiquark respectively, and  $\epsilon^{\lambda'}_{\mu}$ is the polarization
vector of the emitted photon.

Using the fact that the contribution of the second term is the same as that of the first term (except that $e_q\neq e_{\overline{Q}}$), we rewrite above
equation in terms of the electronic charge, $e$ as,
\begin{equation}\label{16}
 M_{fi}=-ie\int \frac{d^4 q}{(2\pi)^4}Tr[\overline{\Psi}_{P}(P', q'){\not}\epsilon'\Psi_{V}(P,q)
S_F^{-1}(-p_2)].
\end{equation}

Now, we reduce the above equation to the effective 3D form by integrating over the over the longitudinal component, $Md\sigma$. This can be expressed as,

\begin{equation}\label{17}
 M_{fi}=-ie\int \frac{d^3 \hat q}{(2\pi)^3}\int\frac{iMd\sigma}{(2\pi )} Tr[\Gamma_P(\hat q')S_F(p'_1) {\not}\epsilon' S_F(p_1) \Gamma_V(\hat q) S_F(-p_2)].
\end{equation}

To calculate $M_{fi}$, we express the propagators, $S_F$ as,
\begin{eqnarray}\label{18}
 &&\nonumber S_F(p_1)=\frac{\Lambda_1^+(\hat q)}{M\sigma+\widehat m_1M-\omega_1+i\epsilon}
 +  \frac{\Lambda_1^-(\hat q)}{M\sigma+\widehat m_1M+\omega_1-i\epsilon},\\&&
\nonumber S_F(-p_2)=\frac{-\Lambda_2^+(\hat q)}{-M\sigma+\widehat m_2M-\omega_2+i\epsilon}
 + \frac{-\Lambda_2^-(\hat q)}{-M\sigma+\widehat m_2M+\omega_2-i\epsilon},\\&&
 S_F(p_1')=\frac{\Lambda_1^+(\hat q')}{M\sigma'+\widehat m_1(-E')-\omega_1'+i\epsilon}
 +  \frac{\Lambda_1^-(\hat q)'}{M\sigma'+\widehat m_1(-E')+\omega_1'-i\epsilon}
 \end{eqnarray}

Here we wish to mention that in transitions involving single photon decays, such as $V\rightarrow P+\gamma$, the
process requires calculation of triangle quark-loop diagram, which involves
two hadron-quark vertices that we attempt in the $4\times 4$ representation of BSE. We now put the propagators into Eq.(\ref{17}), and multiplying this equation from the left by the relation, $\frac{\not P}{M}\frac{\not P}{M}=-1= \frac{\not P}{M}(\Lambda^+_2(\hat{q}')+\Lambda^-_2(\hat{q}'))$ \cite{wang06}, and making use of Eq.(\ref{13}), where $\alpha <1$,  the transition amplitude can be expressed as,

\begin{eqnarray}\label{19}
&&\nonumber M_{fi}=i\int \frac{d^3\hat{q}}{(2\pi)^3} [\Omega_1+\Omega_2+\Omega_3+\Omega_4];\\&&
\nonumber \Omega_1=\int\frac{d\sigma}{(2\pi )}\frac{i}{M^3}  Tr\bigg[ \frac{-{\not}P\Lambda_2^+(\hat q')\Gamma_P(\hat q')\Lambda_1^+(\hat q'){\not}\epsilon'\Lambda_1^+(\hat q)\Gamma_V(\hat q)\Lambda_2^+(\hat q)}{[\sigma-(-\alpha+\widehat m_1\frac{E'}{M}+\frac{\omega'_1}{M})][\sigma-(-\widehat m_1+\frac{\omega_1}{M})][\sigma-(\widehat m_2-\frac{\omega_2}{M})]} \bigg]\\&&
\nonumber \Omega_2=\int\frac{d\sigma}{(2\pi )}\frac{i}{M^3}  Tr\bigg[ \frac{-{\not}P\Lambda_2^+(\hat q')\Gamma_P(\hat q')\Lambda_1^+(\hat q'){\not}\epsilon'\Lambda_1^-(\hat q)\Gamma_V(\hat q)\Lambda_2^-(\hat q)}{[\sigma-(-\alpha+\widehat m_1\frac{E'}{M}+\frac{\omega'_1}{M})][\sigma-(-\widehat m_1-\frac{\omega_1}{M})][\sigma-(\widehat m_2+\frac{\omega_2}{M})]}\bigg];\\&&
\nonumber \Omega_3=\int\frac{d\sigma}{(2\pi )}\frac{i}{M^3}  Tr\bigg[\frac{ -{\not}P\Lambda_2^-(\hat q')\Gamma_P(\hat q')\Lambda_1^-(\hat q'){\not}\epsilon'\Lambda_1^+(\hat q)\Gamma_V(\hat q)\Lambda_2^+(\hat q)}{[\sigma-(-\alpha+\widehat m_1\frac{E'}{M}-\frac{\omega'_1}{M})][\sigma-(-\widehat m_1+\frac{\omega_1}{M})][\sigma-(\widehat m_2-\frac{\omega_2}{M})]}\bigg]\\&&
\Omega_4=\int\frac{d\sigma}{(2\pi )}\frac{i}{M^3}  Tr\bigg[\frac{-{\not}P\Lambda_2^+(\hat q')\Gamma_P(\hat q')\Lambda_1^-(\hat q'){\not}\epsilon'\Lambda_1^-(\hat q)\Gamma_V(\hat q)\Lambda_2^-(\hat q)}{[\sigma-(-\alpha+\widehat m_1\frac{E'}{M}-\frac{\omega'_1}{M})][\sigma-(-\widehat m_1-\frac{\omega_1}{M})][\sigma-(\widehat m_2+\frac{\omega_2}{M})]}\bigg],
\end{eqnarray}

where the rest of the terms are anticipated to be zero on account of 3D Salpeter equations. The contour integrations over $Md\sigma$ are performed over each of the four terms taking into account the pole positions in the complex $\sigma$ plane:

\begin{eqnarray}
&&\nonumber \sigma^{\pm}_{3} =-\alpha+\hat{m}_{1}\frac{E'}{M}\mp \frac{\omega'_{1}}{M}\pm i\epsilon\\&&
\nonumber \sigma^{\pm}_{1} =-\hat{m}_{1}\mp \frac{\omega_{1}}{M}\pm i\epsilon\\&&
\sigma^{\pm}_{2} = \hat{m}_{2} \mp \frac{\omega_2}{M}\pm i\epsilon.
\end{eqnarray}

\begin{figure}[h!]
 \centering
 \includegraphics[width=12cm,height=6cm]{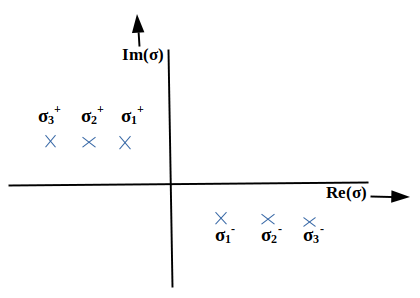}
 \caption{Pole positions in the complex $\sigma$- plane}
 \label{fig:2}
\end{figure}

In Eq.(21)), the contour integral over each of the four terms can be performed by closing the contour either above or below the real axis in the complex $\sigma$- plane with pole positions displayed in Fig.\ref{fig:2}. It can be verified that the results of each of these four integrals, $\Omega_1,...,\Omega_4$ whether we close the contour above or below the real $\sigma$-axis comes out to be the same, thereby validating the correctness of the formalism employed. These results of integrals over $Md\sigma$ in $\Omega_1,...,\Omega_4$, are given as, $\alpha_1,..., \alpha_4$ in Eqs.(23-24).

This leads to the expression for effective 3D form of transition amplitude, $M_{fi}$ under Covariant Instantaneous Ansatz as,

\begin{multline}\label{7f}
 M_{fi}=-ie\int \frac{d^3 \hat q}{(2\pi)^3} \frac{1}{M^2}Tr\bigg[ \alpha_1 {\not}P\overline{\psi}_P^{++}(\hat q'){\not}\epsilon'\psi_V^{++}(\hat q)
 + \alpha_2 {\not}P\overline{\psi}_P^{++}(\hat q'){\not}\epsilon'\psi_V^{--}(\hat q)\\
 +\alpha_3 {\not}P\overline{\psi}_P^{--}(\hat q'){\not}\epsilon'\psi_V^{++}(\hat q)
 + \alpha_4 {\not}P\overline{\psi}_P^{--}(\hat q'){\not}\epsilon'\psi_V^{--}(\hat q)\bigg]
 \end{multline}

where,

\begin{align}
 \nonumber \alpha_1&=\frac{[-E'-\omega'_1-\omega'_2]}{[\alpha-\widehat m_1\frac{E'}{M}+\widehat m_2-\frac{1}{M}(\omega'_1+\omega_2)]}\\
 \nonumber \alpha_2&=\frac{-[-E'-\omega'_1-\omega'_2]}{[\alpha-\widehat m_1(\frac{E'}{M}-1)-\frac{1}{M}(\omega_1+\omega'_1)]}\\
 \nonumber \alpha_3&=\frac{[-E'+\omega'_1+\omega'_2]}{[\alpha-\widehat m_1(\frac{E'}{M}-1)+\frac{1}{M}(\omega_1+\omega'_1)]}\\
  \alpha_4&=\frac{-[-E'+\omega'_1+\omega'_2]}{[\alpha-\widehat m_1\frac{E'}{M}+\widehat m_2+\frac{1}{M}(\omega'_1+\omega_2)]},
\end{align}

and the projected wave functions, $\psi^{\pm \pm}$ being taken from the 3D Salpeter equations \cite{eshete19} derived earlier, which for initial meson in internal variable $\hat{q}$ are given in Section 2. The Salpeter equations in $\hat{q}'$ involve $-E'=\frac{P.P'}{M}$, which is the projection of $P'$ along the direction of initial momentum, $P$.

The factors $(M\pm \omega_1 \pm \omega_2)$ that were also present in the numerators of $\alpha$'s in Eq.(23) as a result of the first two Salpeter equations in variable, $\hat{q}$ in Eqs. (24), get cancelled from the corresponding factors (in denominator) resulting from contour integrals over $Md\sigma$. And as mentioned above it can be verified that the results of each of the four integrals, $\Omega_1,...,\Omega_4$, whether we close the contour above or below the real $\sigma-$ axis comes out to be the same.

Thus, we have given a generalized method for handling quark-triangle diagrams with two hadron -quark vertices in the framework of $4\times 4$ BSE under Covariant Instantaneous Ansatz, by expressing the transition amplitude, $M_{fi}$ (Eq.(21-22)) as a linear superposition of terms involving all possible combinations of $++$, and $--$ components of Salpeter wave functions of final and initial hadrons through not only the $++++$, and $----$ terms but also terms like $++--$, and $--++$, with each of the four terms being associated with a coefficient, $\alpha_i (i=1,...,4)$, which is the result of pole integration in the complex $\sigma$-plane, with pole positions in Eq.(20) (shown in Fig.2). This superposition of all possible terms in Eq.(21-22) should be a feature of relativistic frameworks.

Now, to calculate the process, we need the 4D BS wave functions for vector and pseudoscalar mesons. We again start with the general 4D decomposition of BS
wave functions \cite{smith69}. Using 3D decomposition under Covariant Instantaneous Ansatz, the wave function of vector mesons of dimensionality, $M$
can be
written as \cite{bhatnagar18,hluf16}:

\begin{equation}\label{wf5}
 \psi^V(\hat q)=iM{\not}\epsilon\chi_1(\hat q)+{\not}\epsilon{\not}P\chi_2(\hat q)+[{\not}\epsilon{\not}\hat q-\hat q.\epsilon]\chi_3(\hat q)
 -i[{\not}P{\not}\epsilon{\not}\hat q+\hat q.\epsilon{\not}P]\frac{1}{M}\chi_4(\hat q)+(\hat q.\epsilon)\chi_5(\hat q)-i
 \hat q.\epsilon\frac{{\not}P}{M}\chi_6(\hat q),
\end{equation}

where $\epsilon^{\lambda}$ is the vector meson polarization vector. Similarly for a pseudoscalar meson, the 3D wave function with dimensionality $M$ can
be written as,
\begin{equation}\label{wf2}
 \psi^P(\hat q)=N_{P}[M\phi_1(\hat q)-i{\not}P\phi_2(\hat q)+i{\not}\hat q\phi_3(\hat q)+\frac{{\not}P{\not}\hat q}{M}\phi_4(\hat q)]\gamma_5.
\end{equation}

We wish to mention that in our previous works \cite{hluf16,bhatnagar18,eshete19}, we had calculated the mass spectrum of vector, pseudoscalar and scalar mesons, by using their full Dirac structure of wave functions in Eq.(25), Eq.(26) and Eq.(41) respectively, into the 3D Salpeter equations in Eq.(2), and obtained the coupled Salpeter equations in the amplitudes of various Dirac structures, which were then decoupled using heavy-quark approximation, and mass spectral equations were obtained in an approximate harmonic oscillator basis, which were  used to calculate not only the mass spectrum but also to analytically derive the algebraic forms of wave functions \cite{hluf16,bhatnagar18} in Eq.(29) for both pseudoscalar and vector quarkonia, and in Eq.(43) for scalar mesons\cite{eshete19} in an approximate harmonic basis. And these algebraic wave functions were used to calculate various transitions \cite{hluf16,bhatnagar18,eshete19}, by fixing the parameters of the model to the mass spectrum. Also the plots of these wave functions were studied in details in these works.  And it is these very analytic forms of wave functions in Eqs,(29) and Eq.(43) that we are now using to calculate the M1 and E1 transitions in this work.
\bigskip

However, we further wish to mention that in some of the recent works \cite{alkofer}, it was noticed that among all Dirac covariants in structure of hadronic BS wave function, some covariants contribute much more than others in calculation of hadronic observables. This led us to develop a naive power counting rule in \cite{bhatnagar06,bhatnagar09, bhatnagar14} by which one could classify various Dirac structures as leading and sub-leading. Thus, in our framework \cite{bhatnagar06,bhatnagar09,bhatnagar14}, we had shown that in case of pseudoscalar mesons, the Dirac structures associated with amplitudes $\phi_1$ and $\phi_2$  are leading, while the structures associated with $\phi_3$ and $\phi_4$  are sub-leading. And in various calculations  \cite{bhatnagar06,bhatnagar09,bhatnagar14}, it was shown that the Dirac structure associated with $\phi_1$ (i.e., $\gamma_5$) is dominant. We wish to point out that \cite{munczek,sauli} have also shown that $\phi_1$ (associated with $\gamma_5$) is the most dominant amplitude for not only ground state pseudoscalar mesons, but also their excited states, and this is more true for heavy mesons.
\bigskip

A similar behaviour was observed in case of vector mesons\cite{bhatnagar06,bhatnagar14}, where structures associated with $\chi_1$ and $\chi_2$ are leading, while those associated with $\chi_3,..., \chi_6$ are subleading, and among the leading Dirac structures, the structure associated with $\chi_1$ (i.e. $i\gamma.\epsilon$) is the most dominant. These dominant Dirac structures contribute nearly $80-90\%$ to calculation of any  meson observable, and their contribution \cite{alkofer,bhatnagar06,bhatnagar09, bhatnagar14} increases with increase in meson mass.

Thus, to simplify algebra, we make use of the most dominant Dirac structures for both vector ($iM\gamma.\epsilon$) and pseudoscalar ($M\gamma_5$) mesons, while for radial parts of the wave functions, $\phi_P(\hat q')$, and $\phi_V(\hat q)$, we take their structures as in Eqs.(29),(43), that were obtained through solutions of the mass spectral equations using the full Dirac wave functions in Eqs. (25)-(26). This might lead to the loss of terms involving $\not{P}$, $\not{q}$ etc. that come from the subleading Dirac structures, but their contribution is negligible in comparison to the contribution from the leading Dirac structures used in this work.

Thus, the 4D Bethe-Salpeter wave functions of heavy-light pseudoscalar and
vector quarkonia are taken as,

\begin{equation}\label{wf44}
\begin{split}
  \psi_P(\hat q)=N_P(M'\gamma_5)\\
 \psi_V(\hat q)=N_V(iM{\not}\epsilon)\phi_V(\hat q),
 \end{split}
\end{equation}

Such dominant Dirac structures, $\Gamma =\gamma_5$ (for $0^{-+}$), $\gamma_{\mu}$ (for $1^{--}$), and $1$ (for $0^{++}$) have also been used recently in lattice calculations of radiative decays in \cite{dudek06} recently.

The 4D Bethe-Salpeter normalizers obtained through current conservation condition are:

\begin{equation}
\begin{split}
N_P^{-2}=4\hat m_1 \widehat m_2 M'^2\frac{1}{m_1}\int \frac{d^3\widehat{q}'}{(2\pi)^3}\phi_P^2(\widehat{q}'),\\
N_V^{-2}=4\hat m_1 \widehat m_2 M^2\frac{1}{m_1}\int \frac{d^3\widehat{q}}{(2\pi)^3}\phi_V^2(\widehat{q})
\end{split}
\end{equation}

The 3D wave functions of ground and excited states of pseudoscalar $0^{-+}$ and vector $1^{--}$ quarkonia are \cite{eshete19},
\begin{equation}\label{25}
\begin{split}
   \phi_{P,V}(1S,\hat q)&=\frac{1}{\pi^{3/4}}\frac{1}{\beta_{P,V}^{3/2}}e^{-\frac{\hat q^2}{2\beta_{P,V}^2}}\\
 \phi_{P,V}(2S,\hat q)&=\sqrt{\frac{3}{2}}\frac{1}{\pi^{3/4}}\frac{1}{\beta_{P,V}^{3/2}}
  \bigg(1-\frac{2\hat q^2}{3\beta_{P,V}^2}\bigg)e^{-\frac{\hat q^2}{2\beta_{P,V}^2}}\\
  \phi_V(1D,\hat q)&=\sqrt{\frac{4}{15}}\frac{1}{\pi^{3/4}}\frac{1}{\beta_V^{7/2}}\hat q^2e^{-\frac{\hat q^2}{2\beta_V^2}}\\
\phi_{P,V}(3S,\hat q)&=\sqrt{\frac{15}{8}}\frac{1}{\pi^{3/4}}\frac{1}{\beta_{P,V}^{3/2}}
     \bigg(1-\frac{4\hat q^2}{3\beta_{P,V}^2}+\frac{4\hat q^4}{15\beta_{P,V}^4}\bigg)e^{-\frac{\hat q^2}{2\beta_{P,V}^2}},
 \end{split}
\end{equation}
where the inverse range parameters are
\begin{equation}
   \beta_{P,V}=\bigg(\frac{\frac{1}{2}\omega^2_{q\bar q}(m_1+m_2)}
  {\sqrt{1+8\widehat m_1\widehat m_2A_0(N+\frac{3}{2})}}\bigg)^{1/4}\\
\end{equation}
The $++$ and $--$ components of the B.S. wave function for pseudoscalar meson are \cite{hluf16,wang06}:

\begin{equation}\label{c1}
 \psi_P^{\pm\pm}(\hat q')= \Lambda^{\pm}_{1}(\hat q')\frac{{\not}P}{M}\psi_P(\hat q')
 \frac{{\not}P}{M}\Lambda^{\pm}_{2}(\hat q')
\end{equation}

Substituting the 4D BS wave function of pseudoscalar meson, the $++$ and $--$ components of the 4D BS wave function of pseudoscalar meson can be obtained using Eq.(\ref{c1}) as given in Eq.(\ref{d5}) of Appendix \ref{A1}. The corresponding adjoint wave functions are given in Eq.(\ref{d6}) of Appendix \ref{A1}.

Whereas, the positive and negative energy components of the vector meson wave function are
\begin{equation}\label{c2}
 \psi_V^{\pm\pm}(\hat q)= \Lambda^\pm_{1}(\hat q)\frac{{\not}P}{M}\psi_V(\hat q)
 \frac{{\not}P}{M}\Lambda^\pm_{2}(\hat q)
\end{equation}

Following the same steps as in Eq.(\ref{d71}), we obtain the $++$ and $--$ components of the 4D BS wave function of vector meson through Eq.(\ref{c2}). These components of vector meson wave function are given in Eq.(\ref{d71}), and their corresponding adjoint wave functions are given in Eq.(\ref{d72}) of Appendix \ref{A1}.

We now calculate the individual terms, ${\not}P\overline{\psi}^{++}_{P}(\hat q'){\not}\epsilon'\Psi^{++}_{V}(\hat q)$,  ${\not}P\overline{\psi}^{++}_{P}(\hat q'){\not}\epsilon'\Psi^{--}_{V}(\hat q)$,
${\not}P\overline{\psi}^{--}_{P}(\hat q'){\not}\epsilon'\Psi^{++}_{V}(\hat q)$\\, and ${\not}P\overline{\psi}^{--}_{P}(\hat q'){\not}\epsilon'\Psi^{--}_{V}(\hat q)$ in the transition amplitude, $M_{fi}$. These terms are given in Eqs.(\ref{vp1}-\ref{vp4}) of Appendix \ref{A1}.

Here, it is to be mentioned that, the transverse component of internal momentum of the pseudoscalar meson can be expressed as, $\hat q'=\hat q+ \hat{m}_2\hat{P}'$,as in Eq.(12), where  $\hat{m}_2$ act as momentum partitioning parameters. Now squaring both sides of Eq.(12) that connects $\hat{q}'$, with $\hat{q}$, making use of the fact that $\hat{P}'$, and $\hat{q}$ are both transverse to the initial hadron momentum, and $|\hat{P}'|=|\overrightarrow{P}|=\frac{M^2-M'^2}{2M}$, we can express the relationship between $\hat{q}'^2$, and $\hat{q}^2$ as,

\begin{equation}
\hat{q}'^2=\hat{q}^2+2\hat{m}_2 \frac{(M^2-M'^2)}{2M}|\hat{q}|+\hat{m}_2^2\frac{(M^2-M'^2)^2}{4M^2}
\end{equation}

where,$ |\hat{q}|$ is the length of the effective 3-D vector, $\hat{q}$.  The transition amplitude, $M_{fi}$ is expressed as,

\begin{equation}
M_{fi}=F_{VP}~\epsilon_{\mu\nu\alpha\beta}~P_\mu \epsilon^{\lambda'}_\nu \epsilon^{\lambda}_\alpha P'_\beta ,
\end{equation}
where the antisymmetric tensor, $\epsilon_{\mu\nu\alpha\beta}$ ensures its gauge invariance. Here, $F_{VP}$ is the transition form factor for $V\rightarrow P\gamma$, with expression,

 \begin{eqnarray}\label{11f}
&&\nonumber F_{VP}=-eN_PN_V\frac{M'}{M^3} \int \frac{d^3 \hat q}{(2\pi)^3} \frac{\phi_{P}(\hat q')\phi_{V}(\hat q)}{16\omega_1\omega_2\omega'_1\omega'_2} \bigg[T_1-T_2\frac{M^2-M'^2}{2MM'^2}|\hat{q}|\bigg];\\&&
\nonumber T_1=4(\alpha_1+\alpha_2+\alpha_3+\alpha_4)M^2(m_1-m_2)\hat{m}_2(-\hat{q}^2-\omega_1\omega_2-m_1m_2)\\&&
\nonumber -4(\alpha_1-\alpha_2-\alpha_3+\alpha_4)M^2\hat{m}_2(\omega_1m_2+\omega_2m_1)(\omega_1'+\omega_2')\\&&
\nonumber T_2=4(\alpha_1+\alpha_2+\alpha_3+\alpha_4)M^2(m_1-m_2)\bigg(\omega_1'\omega_2'-\omega_1\omega_2-M(\omega_1'+\omega_2')\hat{m}_2\frac{M^2+M'^2}{M^2}\bigg)\\&&
\nonumber -4(\alpha_1-\alpha_2-\alpha_3+\alpha_4)M^2\bigg(\omega_1'm_2+\omega_2'm_1)(\omega_1+\omega_2)+(\omega_1m_2+\omega_2m_1)(\omega_1'+\omega_2')\bigg)\\&&
+4(\alpha_1-\alpha_2+\alpha_3-\alpha_4)M(m_1-m_2)\hat{m}_2(\omega_1+\omega_2)\frac{M^2+M'^2}{2}
 \end{eqnarray}

The above expression corresponds to $F_{VP}(k^2=0)$, that corresponds to emission of a real photon. However, since in this work, we were mainly interested in calculation of decay widths for various transitions, detailed calculations of $F_{VP}(k^2)$ on lines of \cite{vary18,karmanov15} will be relegated to s separate paper.
Now we proceed to calculate the decay widths for the process $V->P\gamma$, which corresponds to emission of a real photon, for which we need, $F_{VP}(k^2=0)$ given above. The kinematical relation connecting $\hat{q}'^2$, with $\hat{q}^2$, is given in Eq.(34). To calculate the decay widths, we need to calculate the spin averaged amplitude square, $|\overline{M}_{fi}|^2$, where
$|\overline{M}_{fi}|^2=\frac{1}{2j+1}\sum_{\lambda,\lambda'}|{M}_{fi}|^2$, where we average over the initial polarization states $\lambda$ of V-meson, and sum over
the final polarization $\lambda'$ of photon. We make use of the normalizations,
$\Sigma_{\lambda} \epsilon_{\mu}^{\lambda}\epsilon_{\nu}^{\lambda}=\frac{1}{3}(\delta_{\mu\nu}+\frac{P_{\mu}P_{\nu}}{M^2})$ for vector meson, and
$\Sigma_{\lambda'} \epsilon_{\mu}^{\lambda'}\epsilon_{\nu}^{\lambda'}=\delta_{\mu\nu}$, for the emitted photon, with $M_{fi}$ taken from Eq.(34).

The spin-averaged amplitude square of the process, obtained after dividing by the total spin states $(2j+1)$  of the initial vector meson can be
obtained as

\begin{equation}
 |\overline{M_{fi}}|^2=-\frac{2e^2}{3}[M^2M'^2-(P.P')^2]~|F_{VP}(0)|^2
\end{equation}

In the above equation, we evaluate $P.P'=-ME'$ in the rest frame of initial vector meson, where $E'=\sqrt{\overrightarrow{P}'^2+M'^2}$ is the energy of the final pseudoscalar meson, giving,$P.P'=-(\frac{M^2+M'^2}{2})$. Thus, $|\overline{M}_{fi}|^2$ can be expressed as,

\begin{equation}
|\overline{M}_{fi}|^2=\frac{2}{3}e^2\frac{(M^2-M'^2)^2}{4}|F_{VP}(0)|^2.
\end{equation}

The decay width of the process ($V\rightarrow P\gamma$) in the rest frame of the initial vector meson is expressed as
\begin{equation}
 \Gamma_{V\rightarrow P\gamma}=\frac{|\overline{M}_{fi}|^2}{8\pi M^2}|\overrightarrow{P'}|,
\end{equation}

where we make use of the fact that modulus of the momentum of the emitted pseudoscalar meson can be expressed in terms of masses of particles as,
$|\overrightarrow{P'}|=|\overrightarrow{k}|=\omega_k=\frac{1}{2M} (M^2-M'^2)$, where, $\omega_k$ is the kinematically allowed energy of the emitted photon. Thus, $\Gamma$ in turn can be expressed as:

\begin{equation}
\Gamma=\frac{\alpha_{e.m.}}{3}|F_{VP}|^2\omega_k^3.
\end{equation}

We now calculate the radiative decay widths for the process, $V\rightarrow S+\gamma$ in the next section.

\section{Radiative decays of heavy-light quarkonia through $V\rightarrow S\gamma$}
E1 transitions always involve excited states. The scattering amplitude of the decay process $V\rightarrow S\gamma$ can be written as
\begin{multline}\label{7f1}
 M_{fi}=-ie\int \frac{d^3 \hat q}{(2\pi)^3} \frac{1}{M^2}Tr\bigg[ \alpha_1 {\not}P\overline{\psi}_S^{++}(\hat q'){\not}\epsilon'\psi_V^{++}(\hat q)
 + \alpha_2 {\not}P\overline{\psi}_S^{++}(\hat q'){\not}\epsilon'\psi_V^{--}(\hat q)\\
 +\alpha_3 {\not}P\overline{\psi}_S^{--}(\hat q'){\not}\epsilon'\psi_V^{++}(\hat q)
 + \alpha_4 {\not}P\overline{\psi}_S^{--}(\hat q'){\not}\epsilon'\psi_V^{--}(\hat q)\bigg]
 \end{multline}

After the 3D reduction of the 4D BS wave function of scalar meson under CIA, we express the 3D BS wave function with dimensionality $M$ as

\begin{equation}
\psi_S(\hat q)=N_S[Mf_1(\hat{q})+i{\not}P f_2(\hat{q})-i{\not}\hat{q}f_3(\hat{q})+2\frac{{\not}P{\not}\hat{q}}{M}f_4(\hat{q})].
\end{equation}
Making use of the fact that the most leading Dirac structure in scalar meson BS wave function is $MI$ ($I$ being the unit $4\times4$ unit matrix), and making
use of \cite{bhatnagar18}, we express the 3D scalar meson BS wave function as,

\begin{equation}\label{sc1}
 \psi_S(\hat q)=N_S(M')\phi_S(\hat q'),
\end{equation}

where $\phi_S(\hat{q})$ is the spatial part of this wave function, whose analytic form is obtained by solving the 3D mass spectral equations for scalar mesons, given in \cite{bhatnagar18} are

\begin{equation}\label{wv1}
\begin{split}
 \phi_S(1P,\hat q)&=\sqrt{\frac{2}{3}}\frac{1}{\pi^{3/4}}\frac{1}{\beta_S^{5/2}} \hat qe^{-\frac{\hat q^2}{2\beta_S^2}}\\
 \phi_S(2P,\hat q)&=\sqrt{\frac{5}{3}}\frac{1}{\pi^{3/4}}\frac{1}{\beta_S^{5/2}}
  \hat q\bigg(1-\frac{2\hat q^2}{5\beta_S^2}\bigg)e^{-\frac{\hat q^2}{2\beta_S^2}}\\
    \phi_S(3P,\hat q)&=\sqrt{\frac{35}{12}}\frac{1}{\pi^{3/4}}\frac{1}{\beta_S^{5/2}}
 \hat q\bigg(1-\frac{4\hat q^2}{5\beta_S^2}+\frac{4\hat q^4}{35\beta_S^4}\bigg)e^{-\frac{\hat q^2}{2\beta_S^2}}\\
   \phi_S(4P,\hat q)&=\sqrt{\frac{35}{8}}\frac{1}{\pi^{3/4}}\frac{1}{\beta_S^{5/2}}
 \hat q\bigg(1-\frac{6\hat q^2}{5\beta_S^2}+\frac{12\hat q^4}{35\beta_S^4}-\frac{8\hat q^6}{315\beta_S^6}\bigg)e^{-\frac{\hat q^2}{2\beta_S^2}},
\end{split}
\end{equation}

The 4D BS normalizer of scalar meson, $N_S$, can be obtained by solving the current conservation conditions, and is
expressed as,
\begin{equation}
 N_S^{-2}=4\hat m_1 \widehat m_2 M'^2\frac{1}{m_1}\int \frac{d^3\widehat{q}}{(2\pi)^3}\phi_S^2(\widehat{q}').\\
\end{equation}

We now obtain the $++$ and $--$ components of the scalar meson wave function through Eq.(\ref{c1}) as given in Eq.(\ref{e78}) with the corresponding adjoint wave functions in Eq.(\ref{e88}) of Appendix \ref{A1}.
The expressions for $++++$, $++--$, $--++$, and $----$ terms of the scattering amplitude in Eq.(\ref{7f}) is  relegated to Appendix \ref{A2}.

We first evaluate trace over the gamma matrices in Eq.(40). We make use of the fact that $\hat{q}'=\hat{q}+\hat{m}_2 \hat{P}'$, where, $\hat{P}'=P'-\frac{P'.P}{P^2}P$. We combine various terms, and further make use of the fact that, for initial vector meson, $P.\epsilon^{\lambda}=0$, and in its rest frame, $P'.\epsilon^{\lambda'}=0$ (where $\epsilon^{\lambda'}$ is the photon polarization vector). Due to this, we express $\hat{P}'.\epsilon= P'.\epsilon$, and $\hat{P}'.\epsilon'=\beta P.\epsilon'$, where $\beta=-\frac{P'.P}{P^2}=-\frac{M^2+M'^2}{2M^2}$ from Eq.(13). We can then express the invariant matrix element, $M_{fi}$ as,

\begin{equation}\label{9f2}
 M_{fi}=-ie N_S N_V\frac{1}{M^2} \int \frac{d^3 \hat q}{(2\pi)^3} \frac{\phi_{S}(\hat q')\phi_{V}(\hat q)}{16\omega_1\omega_2\omega'_1\omega'_2} [\Theta_1(\epsilon^{\lambda'}.\epsilon^{\lambda})\\
 +\Theta_2\beta(\epsilon^{\lambda'}.P)(\epsilon^{\lambda}.P')],
 \end{equation}

 \begin{eqnarray}
  &&\nonumber  \Theta_1= 4M^3\bigg( (\alpha_1-\alpha_2+\alpha_3-\alpha_4)(\omega'_1\omega'_2-m_1m_2+\hat q'^2)(\omega_1m_2+m_1\omega_2) \\&&
 \nonumber
 + (\alpha_1+\alpha_2-\alpha_3-\alpha_4)(\omega'_1m_2-m_1\omega'_2)(\omega_1\omega_2+m_1m_2+\hat q^2)
 + [ (\alpha_1+\alpha_2-\alpha_3-\alpha_4)(m_1-m_2)(\omega'_1+\omega'_2) \\&&
 \nonumber
 -(\alpha_1-\alpha_2+\alpha_3-\alpha_4)(m_1+m_2)(\omega_1-\omega_2) ] (\hat q^2+\widehat m_2\frac{M^2-M'^2}{2M}|\hat{q}|) \bigg)\\&&
\nonumber \Theta_2=\frac{16M^5\hat{q}^2}{(M^2-M'^2)^2}\bigg((\alpha_1+\alpha_2-\alpha_3-\alpha_4) [-(\omega'_1m_2-m_1\omega'_2)
+m_2(\omega'_1+\omega'_2)]
 +(\alpha_1-\alpha_2+\alpha_3-\alpha_4)\omega_2 (m_1+m_2) \bigg)\\&&
 +\frac{8M^4|\hat{q}|}{(M^2-M'^2)} \widehat m_2 \bigg(-(\alpha_1+\alpha_2-\alpha_3-\alpha_4)(m_1-m_2)(\omega'_1+\omega'_2)
+(\alpha_1-\alpha_2+\alpha_3-\alpha_4)(m_1+m_2)(\omega_1+\omega_2) \bigg).
\end{eqnarray}
\bigskip

After carrying out the integrals $d^3\hat{q}$ over $\Theta_1$, and $\Theta_2$, in Eq.(45), we can express the amplitude, $M_{fi}$ as,

\begin{eqnarray}
&&\nonumber M_{fi}=S_1(\epsilon^{\lambda'}.\epsilon^{\lambda})+S_2\beta(\epsilon^{\lambda'}.P)(\epsilon^{\lambda}.P'),\\&&
\nonumber S_1=-ie N_SN_V\frac{1}{M^2} \int \frac{d^3 \hat q}{(2\pi)^3} \frac{\phi_{S}(\hat q')\phi_{V}(\hat q)}{16\omega_1\omega_2\omega'_1\omega'_2} \Theta_1,\\&&
S_2=-ieN_SN_V\frac{1}{M^2} \int \frac{d^3 \hat q}{(2\pi)^3} \frac{\phi_{S}(\hat q')\phi_{V}(\hat q)}{16\omega_1\omega_2\omega'_1\omega'_2}\Theta_2,
\end{eqnarray}

And, $S_1$, and $S_2$ are the form factors. Now, to calculate the decay widths, we need to calculate the spin averaged amplitude square, $|\overline{M}_{fi}|^2$, where
$|\overline{M}_{fi}|^2=\frac{1}{2j+1}\sum_{\lambda,\lambda'}|{M}_{fi}|^2$, where we average over the initial polarization states $\lambda$ of V-meson, and sum over
the final polarization $\lambda'$ of photon. We make use of the normalizations,
$\Sigma_{\lambda} \epsilon_{\mu}^{\lambda}\epsilon_{\nu}^{\lambda}=\frac{1}{3}(\delta_{\mu\nu}+\frac{P_{\mu}P_{\nu}}{M^2})$ for vector meson, and
$\Sigma_{\lambda'} \epsilon_{\mu}^{\lambda'}\epsilon_{\nu}^{\lambda'}=\delta_{\mu\nu}$, for the emitted photon, with $M_{fi}$ taken from the previous
equations. This gives
$\sum_{\lambda'}\sum_{\lambda}|\epsilon^{\lambda'}.\epsilon^{\lambda}|^2 =1$.

The spin-averaged amplitude square of the process can be written as
\begin{equation}
 |\overline{M}_{fi}|^2=\frac{1}{3} \bigg[ |S_1|^2+ \frac{1}{3}\beta^2 [M^2 M'^2-(P'.P)^2]|S_2|^2\bigg],
\end{equation}

where, $\beta^2=\frac{(M^2+M'^2)^2}{4M^2}$. We can write the decay width,

\begin{equation}\label{50}
\Gamma_{V\rightarrow S\gamma}=\frac{|\overline{M}_{fi}|^2}{8\pi M^2}|\overrightarrow{P'}|,
\end{equation}
where we make use of the fact that modulus of the momentum of the emitted pseudoscalar meson can be expressed in terms of masses of particles as,
$|\overrightarrow{P'}|=\frac{1}{2M} (M^2-M'^2)$.

\section{Radiative decays of heavy-light quarkonia through $S\rightarrow V\gamma$}
We proceed to evaluate the process in the same manner as $V \rightarrow S\gamma$, using Fig.\ref{fig:1}, where the initial scalar meson decays into a vector meson and a photon. Drawing analogy from $V\rightarrow P\gamma$, and $V\rightarrow S\gamma$, the effective 3D form of transition amplitude, $M_{fi}$ for $S\rightarrow V\gamma$ under Covariant Instantaneous Ansatz can be expressed as,

\begin{multline}\label{7f3}
 M_{fi}=-ie\int \frac{d^3 \hat q}{(2\pi)^3} \frac{1}{M^2}Tr\bigg[ \alpha_1 {\not}P\overline{\psi}_V^{++}(\hat q'){\not}\epsilon'\psi_S^{++}(\hat q)
 + \alpha_2 {\not}P\overline{\psi}_V^{++}(\hat q'){\not}\epsilon'\psi_S^{--}(\hat q)\\
 +\alpha_3 {\not}P\overline{\psi}_V^{--}(\hat q'){\not}\epsilon'\psi_S^{++}(\hat q)
 + \alpha_4 {\not}P\overline{\psi}_V^{--}(\hat q'){\not}\epsilon'\psi_S^{--}(\hat q)\bigg].
 \end{multline}

The transition amplitude of the $S\rightarrow V\gamma$ process can be obtained as
\begin{equation}\label{9f3}
 M_{fi}=eN_SN_V \frac{M'}{M^3} \int \frac{d^3 \hat q}{(2\pi)^3} \frac{\phi_{S}(\hat q')\phi_{V}(\hat q)}{16\omega_1\omega_2\omega'_1\omega'_2} ~[TR]
\end{equation}
where
\begin{multline}
 [TR] = Tr[ -(\alpha_1-\alpha_2+\alpha_3-\alpha_4)
 iM^3(\omega'_1\omega'_2+m_1m_2)(\omega_1m_2-m_1\omega_2){\not}\epsilon^{\lambda}{\not}\epsilon^{\lambda'}\\
  +(\alpha_1-\alpha_2+\alpha_3-\alpha_4) iM^3(\omega_1m_2-m_1\omega_2){\not}\hat q'{\not}\epsilon^{\lambda}{\not}\hat q'{\not}\epsilon^{\lambda'}\\
 -(\alpha_1+\alpha_2+\alpha_3+\alpha_4) iM^2(m_1+m_2){\not}P{\not}\hat q'{\not}\epsilon^{\lambda}{\not}\hat q'{\not}\epsilon^{\lambda'}{\not}\hat q\\
 -(\alpha_1+\alpha_2-\alpha_3-\alpha_4) iM^3(\omega'_1m_2+m_1\omega'_2)(\omega_1\omega_2-m_1m_2+\hat q^2){\not}\epsilon^{\lambda}{\not}\epsilon^{\lambda'}\\
    -(\alpha_1-\alpha_2+\alpha_3-\alpha_4) iM^3(m_1(\omega_1+\omega_2){\not}\hat q'{\not}\epsilon^{\lambda}{\not}\epsilon^{\lambda'}{\not}\hat q
 +m_2(\omega_1+\omega_2){\not}\epsilon^{\lambda}{\not}\hat q'{\not}\epsilon^{\lambda'}{\not}\hat q)\\
 +(\alpha_1+\alpha_2-\alpha_3-\alpha_4) iM^3(m_1+m_2)(\omega'_1{\not}\hat q'{\not}\epsilon^{\lambda}{\not}\epsilon^{\lambda'}{\not}\hat q -\omega'_2{\not}\epsilon^{\lambda}{\not}\hat q'{\not}\epsilon^{\lambda'}{\not}\hat q ) ]
\end{multline}

We first evaluate trace over the gamma matrices in above equation. We make use of the fact that $\hat{q}'=\hat{q}+\hat{m}_2 \hat{P}'$, where, $\hat{P}'=P'-\frac{P'.P}{P^2}P$. We combine various terms, and further make use of the fact that, for initial scalar meson at rest in its own frame,
$P.\epsilon^{\lambda'}=0$, and  $P'.\epsilon^{\lambda}=0$, where $\epsilon^{\lambda}$ is the polarization vector of emitted vector meson with momentum, $P'$, and  $\epsilon^{\lambda'}$ is the photon polarization vector. Due to this, we express $\hat{P}'.\epsilon= \beta (P.\epsilon)$, and $\hat{P}'.\epsilon'= P'.\epsilon'$, where $\beta=-\frac{P'.P}{P^2}=-\frac{M^2+M'^2}{2M^2}$. We can then express the invariant matrix element, $M_{fi}$ as,

\begin{equation}
 M_{fi}=-ie N_S N_V\frac{1}{M^2} \int \frac{d^3 \hat q}{(2\pi)^3} \frac{\phi_{S}(\hat q')\phi_{V}(\hat q)}{16\omega_1\omega_2\omega'_1\omega'_2} [\Theta'_1(\epsilon^{\lambda'}.\epsilon^{\lambda})\\
 +\Theta'_2\beta (\epsilon^{\lambda}. P)(\epsilon^{\lambda'}.P')],
\end{equation}
where

\begin{eqnarray}
&&\nonumber \Theta_1'=4(\alpha_1-\alpha_2+\alpha_3-\alpha_4)M^3\bigg((-m_2\omega_1+m_1\omega_2)(m_1m_2+\omega_1'\omega_2'+\hat{q}'^2)\\&&
\nonumber +\frac{1}{2M}\bigg[(\hat{m}_2(M^2-M'^2)|\hat{q}|+2M\hat{q}^2)(m_1-m_2)(\omega_1+\omega_2)\bigg]\bigg)-4M^3(\alpha_1+\alpha_2-\alpha_3-\alpha_4)\\&&
\nonumber \bigg((\omega_1\omega_2-m_1m_2+\hat{q}^2)(m_2\omega_1'+m_1\omega_2')+\frac{1}{2M}(\hat{m}_2(M^2-M'^2)|\hat{q}|+2M\hat{q}^2)(m_1+m_2)(\omega_1'+\omega_2')\bigg);\\&&
\nonumber \Theta_2'=(\alpha_1-\alpha_2+\alpha_3-\alpha_4)M^3\bigg(\hat{m}_2^2(m_2\omega_1-m_1\omega_2)-32\frac{M^2}{(M^2-M'^2)^2}(m_1+m_2)\omega_2\hat{q}^2\\&&
\nonumber -16\frac{M\hat{m}_2}{M^2-M'^2}|\hat{q}|(-m_2\omega_1+2m_1\omega_2+m_2\omega_2)\bigg)\\&&
+(\alpha_1+\alpha_2-\alpha_3-\alpha_4)M^3\bigg(-\frac{32M^2}{(M^2-M'^2)^2}(m_1+m_2)\omega_2'\hat{q}^2-16\frac{M\hat{m}_2}{(M^2-M'^2)}|\hat{q}|(m_1+m_2)\omega_2'\bigg)
\end{eqnarray}

Thus, $M_{fi}$ can be expressed as,
\begin{eqnarray}
&&\nonumber M_{fi}=S'_1(\epsilon^{\lambda'}.\epsilon^{\lambda})+S'_2 \beta (\epsilon^{\lambda'}.P')(\epsilon^{\lambda}.P),\\&&
\nonumber S'_1=-ie N_SN_V\frac{1}{M^2} \int \frac{d^3 \hat q}{(2\pi)^3} \frac{\phi_{S}(\hat q')\phi_{V}(\hat q)}{16\omega_1\omega_2\omega'_1\omega'_2} \Theta_1',\\&&
S'_2=-ieN_SN_V\frac{1}{M^2} \int \frac{d^3 \hat q}{(2\pi)^3} \frac{\phi_{S}(\hat q')\phi_{V}(\hat q)}{16\omega_1\omega_2\omega'_1\omega'_2}\Theta_2',
\end{eqnarray}

To calculate the decay widths, we again need to calculate the spin averaged amplitude square, $|\overline{M}_{fi}|^2$, where
$|\overline{M}_{fi}|^2=\sum_{\lambda,\lambda'}|{M}_{fi}|^2$, where we sum over the final polarization states, $\lambda'$ of photon, and  $\lambda$ of V-meson.

The spin averaged amplitude modulus square gives,
\begin{equation}
 |\overline{M}_{fi}|^2= \bigg[ |S'_1|^2+\frac{1}{3}\beta^2[P^2P'^2-(P'.P)^2]|S'_2|^2\bigg].
\end{equation}

The decay widths $\Gamma$ for the process, $S\rightarrow V \gamma$, are given by Eq.(\ref{50}), with $P'$, now the momentum of the emitted vector meson.

\section{Results and Discussion}
We have studied radiative decays of conventional heavy-light quarkonia through M1 and E1 transitions in the framework of
Bethe-Salpeter equation. Such processes involve quark-triangle diagrams, and involve two hardon-quark vertices and are difficult to evaluate in BSE under CIA [14,16]. In this work we have given a generalized method of handling quark triangle diagrams with two hadron-quark vertices  in the framework of $4\times 4$ BSE, by expressing the transition amplitude, $M_{fi}$ (Eq.(21-22)) as a linear superposition of terms involving all possible combinations of $++$, and $--$ components of Salpeter wave functions of final and initial hadrons, through not only the etrms, $++++$, and $----$, but also the terms like, $++--$, and $--++$, with each of the four terms being associated with a coefficient, $\alpha_i (i=1,...,4)$, which is the result of pole integration in the complex $\sigma$-plane, with pole positions in Eq.(20) (shown in Fig.2). This superposition of all possible terms in Eq.(21-22) should be a feature of relativistic frameworks.

Using this generalized expression for $M_{fi}$, in Eq.(21-22), we have evaluated the decay widths for $M1$ transitions,  $^3S_1 \rightarrow ^1S_0 +\gamma$, involving the decays of the ground and excited states of the heavy-light mesons such as, $B_c*, B*, J/\Psi, D*, and D_s*$. Here, we have studied the processes, $nS\rightarrow n'S$ for both $n=n'$, and $n\neq n'$. As regards the $E1$ transitions, we have studied the processes, $^3S_1 \rightarrow ^1P_0 +\gamma$, that involve the decays of $\Psi(2S), B_c*(2S)$, and $ D*(2S)$, and the processes, $^1P_0 \rightarrow ^3S_1 +\gamma$, that involve decays of $\chi_{c0}(1P), B_c(1P)$ and $B_c(2P)$.

We used algebraic forms of 3D Salpeter wave functions obtained through analytic solutions of mass spectral equations in approximate harmonic oscillator basis for ground and excited states of $0^{++},1^{--}$, and $0^{-+}$ heavy-light quarkonia for calculation of their decay widths. The input parameters used by us are: $C_0$= 0.69, $\omega_0$= 0.22 GeV, $\Lambda_{QCD}$= 0.25 GeV, and $A_0$= 0.01, along with the input quark masses $m_u$= 0.30 GeV, $m_s$= 0.43 GeV, $m_c$= 1.49 GeV, and $m_b$= 4.67 GeV., that were obtained by fitting to their mass spectra\cite{eshete19}. We have compared our results with experimental data and other models, and found reasonable agreements.

We get reasonable agreements of our decay widths for M1 transitions, $nS\rightarrow n'S+\gamma$  (with $n'= n$), and also for $n'\neq n$. This can be seen from Table 1, for the transitions,  $J/\Psi(1S) \rightarrow \eta_c(1S)$,  $\Psi(2S)\rightarrow \eta_c(2S)$, and $\Psi(2S)\rightarrow \eta_c(1S)$. Similar agreements of our decay widths for $E1$ transitions are noticed for $nS\rightarrow n'P+\gamma$ , and $nP\rightarrow n'S+\gamma$ for both $n =n'$, and $n\neq n'$, as can be seen from Table 2.

We wish to mention that $F_{VP}(0)$ in Eq.(34) are the electromagnetic coupling constants, $g_{VP\gamma}$. It is seen that our coupling constant, $g_{J/\Psi \eta_c\gamma}=0.745GeV^{-1} (Expt.=0.570\pm 0.110 GeV^{-1}$ \cite{pdg16}, while the coupling constant, $g_{D*D\gamma}=-0.438GeV^{-1}$, which can be compared with experimental data that gives $-0.466GeV^{-1}$ \cite{pdg16}, and $-0.384GeV^{-1}$ \cite{Choi07}. Our $g_{D_s*D_s\gamma}=-0.173GeV^{-1}$ , which is comparable to the RQM model value $0.161GeV^{-1}$ \cite{Choi07}. Our $g_{B_s*B_s\gamma}=-0.4773 GeV^{-1}$ that can be compared with $-0.536$\cite{Choi07} and $-0.657$\cite{prd16}. Similarly, our $g_{B*B\gamma}=-0.764GeV^{-1}$, that can be compared with $-0.749$\cite{Choi07}, and $-0.891$\cite{prd16}. However, these results show that various models have a wide range of variations of coupling constants, $g_{VP\gamma}$ for different transitions.

Similarly we again see a wide range of variations in different models for $M1$ transitions, particularly for decays of $J/\Psi$, and $\Psi(2S)$. Further, our $nS\rightarrow nS$ transitions show a marked decrease as we go from ground to higher excited states, which is in conformity with data and other models. We have also given our predictions for radiative decays of $B_c*(1S), B_c*(2S), B_s*(2S), B*(2S), D*(2S)$, for which data is not yet available, and for $D_s*(1S)$, where PDG \cite{Patrignani19} gives only the upper limit of the decay width. As regards $E1$ transitions, our decay width result for $\Psi(2S)$ is in good agreement with data, but for $\chi_{c0}$ is higher than data, though again there is a lot of variation in results of other models. These results have been obtained using leading Dirac structures in the wave functions of P, V and S mesons, though incorporation of all Dirac structures is expected to give better agreement with data.

The aim of doing this study was to mainly test our analytic forms of wave functions in Eqs.(\ref{25}) and \ref{wv1}, obtained as solutions of mass spectral equations in an approximate harmonic oscillator basis obtained analytically from $4 \times 4$ BSE as a starting point, that has so far given good predictions \cite{eshete19,bhatnagar18, hluf16} not only of the mass spectrum of heavy-light quarkonia, but also their leptonic decays, two-photon, and two gluon decays. The present work would in turn lead to the validation of our approach, which provides a much deeper insight than the purely numerical calculations in $4 \times 4$ BSE approach that are prevalent in the literature.

This work was mainly focused on evaluation of decay widths for $M1$, and $E1$ transitions. A more detailed study on not only the transition form factors of both $M1$, and $E1$ transitions, but also the "static" form factors describing meson-photon interactions through the vertex $M\gamma M$ for various mesons will be relegated to a separate paper.

Acknowledgement: This work was carried out at Chandigarh University, and Addis Ababa University. The authors wish to thank both the institutions for the facilities provided during the course of this work. EG would like to thank SIDA for facilitating his visit to Chandigarh University during the course of this work.

\bigskip
\begin{center}
\begin{tabular}{p{4.5cm} p{3cm} p{3.1cm} p{2.4cm} p{1.5cm} p{1.6cm} }
  \hline\hline
 &BSE-CIA &~~~Expt.  &LFQM&PM  &RQM\\
   \hline
  $\Gamma_{J/\psi(1S_1)\rightarrow\eta_c(1S_0)\gamma}$& 1.7036&1.5793$\pm$0.0112\cite{pdg10}& 1.69$\pm$0.05\cite{Choi07} &1.8\cite{repko07}& 1.050\cite{Ebert03}\\
  $\Gamma_{\psi(2S_1)\rightarrow\eta_c(2S_0)\gamma}$& 0.18204&0.2002$\pm$0.008\cite{pdg14}&   &0.4\cite{repko07}  &\\
  $\Gamma_{\psi(2S_1)\rightarrow\eta_c(1S_0)\gamma}$& 0.9340&0.9724&   &  &\\

  $\Gamma_{B_c^*(1S_1)\rightarrow B_c(1S_0)\gamma}$&  0.0664 &  &  &0.06\cite{Gershtein95}&0.033\cite{Ebert03} \\
  $\Gamma_{B_c^*(2S_1)\rightarrow B_c(2S_0)\gamma}$& 0.0360 &  && 0.01\cite{Gershtein95}  & 0.017\cite{Ebert03}\\

 $\Gamma_{B_s^*(1S_1)\rightarrow B_s(1S_0)\gamma}$& 0.0624 &0.064$\pm$0.016\cite{Patrignani19} &0.068$\pm$0.017\cite{Choi07}  &  & \\
  $\Gamma_{B_s^*(2S_1)\rightarrow B_s(2S_0)\gamma}$&  0.04708 &  &&     &\\

 $\Gamma_{B^*(1S_1)\rightarrow B(1S_0)\gamma}$& 0.1364 &0.13$\pm$0.01\cite{Patrignani19}  &0.13$\pm$0.01\cite{Choi07}  &  & \\
  $\Gamma_{B^*(2S_1)\rightarrow B(2S_0)\gamma}$&  0.1467&   &&     &\\
$\Gamma_{D_s^*(1S_1)\rightarrow D_s(1S_0)\gamma}$& 0.2018& &0.17$\pm$ 0.01 \cite{Patrignani19}  &    &0.213\cite{prd16}\\
  $\Gamma_{D^*(1S_1)\rightarrow D(1S_0)\gamma}$& 1.2843& 1.3344$\pm$0.0072\cite{Patrignani19}&0.90$\pm$0.02\cite{Choi07}  $\  $\ &     & \\
  $\Gamma_{D^*(2S_1)\rightarrow D(2S_0)\gamma}$& 0.1381&   & &   & \\
     \hline\hline
   \end{tabular}
   \\ Table 1. Radiative decay widths of heavy-light mesons (in Kev) for M1 transitions in BSE, along with experimental data  and results of other models.
   \end{center}

\bigskip
\begin{center}
\begin{tabular}{p{2.7cm} p{1.5cm} p{3.3cm} p{2cm} p{1.9cm} p{1.6cm} }
  \hline\hline
 &BSE-CIA &~~~Expt.  &PM  &RQM\\
   \hline
  $\Gamma_{\psi(2S_1)\rightarrow \chi_{c0}(1P_0)\gamma}$ &34.0419 &28.5714$\pm$0.0432\cite{pdg14}  &   &26.3\cite{Ebert03} \\
  $\Gamma_{\psi(3S_1)\rightarrow \chi_{c0}(2P_0)\gamma}$ &62.229                      &  &51.4\cite{chnphys}   &65.7\cite{repko07} \\
  $\Gamma_{\psi(3S_1)\rightarrow \chi_{c0}(1P_0)\gamma}$ &1.4441                                 &  &1.2\cite{chnphys}   &    \\
    $\Gamma_{B_c^*(2S_1)\rightarrow B_c(1P_0)\gamma}$ &10.5249&    &9.6\cite{Fulcher99}&   3.78\cite{Ebert03} \\
 $\Gamma_{D^*(2S_1)\rightarrow D(1P_0)\gamma}$ &1.0214 &   &&    \\
 &&&&\\
 $\Gamma_{ \chi_{c0}(1P_0)\rightarrow J/\psi(1S_1)\gamma} $ &123.803&119.5$\pm 8$\cite{Patrignani19} &&161 \cite{Ebert03}   \\
 $\Gamma_{ \chi_{c0}(2P_0)\rightarrow \psi(2S_1)\gamma} $ &75.229     & &68\cite{chnphys}   &    \\
 $\Gamma_{ \chi_{c0}(2P_0)\rightarrow J/\psi(1S_1)\gamma}$&129.86    & &146\cite{chnphys}& 21 \cite{repko07} \\
 $\Gamma_{ B_c(1P_0)\rightarrow B_c^*(1S_1)\gamma} $ &68.580& &65.3\cite{Gershtein95}& 75.5\cite{Ebert03}   \\
$\Gamma_{ B_c(2P_0)\rightarrow B_c^*(2S_1)\gamma} $ &51.3911& &52.5\cite{Gershtein95}& 34\cite{Ebert03}   \\
     \hline\hline
   \end{tabular}
  \\ Table 2. Radiative decay widths of heavy-light mesons (in KeV.) for E1 transitions, along with experimental data  and results of other models.
   \end{center}

   \appendix
\section{Appendix}
\subsection{Radiative decays through V$\rightarrow$ P$\gamma$} \label{A1}
\renewcommand{\theequation}{\thesection.\arabic{equation}}

Substituting the 4D BS wave function of pseudoscalar meson in Eq.(\ref{wf44}), we obtain the $++$ and $--$ components as
\begin{align}\label{d5}
\nonumber \psi^{++}_P(\hat q')&=\frac{N_P}{4\omega'_1\omega'_2}\frac{M'}{M}\phi_{P}(\hat q')[M(\omega'_1\omega'_2+m_1m_2+\hat q'^2)-i(\omega'_1m_2+m_1\omega'_2){\not}P
  +iM(m_1-m_2){\not}\hat q'
  +(\omega'_1{{\not}P\not}\hat q'-\omega'_2{\not}\hat q'{\not}P)]\gamma_5\\
  \psi^{--}_P(\hat q')&=\frac{N_P}{4\omega'_1\omega'_2}\frac{M'}{M}\phi_{P}(\hat q')[M(\omega'_1\omega'_2+m_1m_2+\hat q'^2)+i(\omega'_1m_2+m_1\omega'_2){\not}P
  +iM(m_1-m_2){\not}\hat q'
  -(\omega'_1{{\not}P\not}\hat q'-\omega'_2{\not}\hat q'{\not}P)]\gamma_5
  \end{align}

The adjoint Bethe-Salpeter wave function of pseudoscalar meson can be obtained by evaluating
$\overline{\psi}^{\pm\pm}_P(\hat q')=\gamma_4(\psi^{\pm\pm}_P(\hat q'))^+\gamma_4$ as
\begin{align}\label{d6}
\nonumber \overline{\psi}^{++}_P(\hat q')&=\frac{N_P}{4\omega'_1\omega'_2}\frac{M'}{M}\phi_{P}(\hat q')[-M(\omega'_1\omega'_2+m_1m_2+\hat q'^2)-i(\omega'_1m_2+m_1\omega'_2){\not}P
  +iM(m_1-m_2){\not}\hat q' -(\omega'_1{\not}\hat q'{\not}P-\omega'_2{\not}P{\not}\hat q')]\gamma_5\\
  \overline{\psi}^{--}_P(\hat q')&=\frac{N_P}{4\omega'_1\omega'_2}\frac{M'}{M}\phi_{P}(\hat q')[-M(\omega'_1\omega'_2+m_1m_2+\hat q'^2)+i(\omega'_1m_2+m_1\omega'_2){\not}P
  +iM(m_1-m_2){\not}\hat q' +(\omega'_1{\not}\hat q'{\not}P-\omega'_2{\not}P{\not}\hat q')]\gamma_5
\end{align}

Following the same steps as in Eq.(\ref{d71}), we obtain the $++$ and $--$ components of vector meson wave function in Eq.(\ref{wf44}) as
\begin{align}\label{d71}
\nonumber \psi^{++}_V(\hat q)=\frac{iN_V}{4\omega_1\omega_2}\phi_{V}(\hat q)[
M(\omega_1\omega_2+m_1m_2){\not}\epsilon-M{\not}\hat q{\not}\epsilon{\not}\hat q
+i(\omega_1m_2+m_1\omega_2){\not}\epsilon{\not}P
-iM(m_1{\not}\epsilon{\not}\hat q+m_2{\not}\hat q{\not}\epsilon)
+(\omega_1{\not}\epsilon{\not}P{\not}\hat q-\omega_2{\not}\hat q{\not}P{\not}\epsilon)]\\
\psi^{--}_V(\hat q)=\frac{iN_V}{4\omega_1\omega_2}\phi_{V}(\hat q)[
M(\omega_1\omega_2+m_1m_2){\not}\epsilon-M{\not}\hat q{\not}\epsilon{\not}\hat q
-i(\omega_1m_2+m_1\omega_2){\not}\epsilon{\not}P
-iM(m_1{\not}\epsilon{\not}\hat q+m_2{\not}\hat q{\not}\epsilon)
-(\omega_1{\not}\epsilon{\not}P{\not}\hat q-\omega_2{\not}\hat q{\not}P{\not}\epsilon)],
\end{align}
where as the adjoint wave functions are
\begin{align}\label{d72}
\nonumber \overline{\psi}^{++}_V(\hat q)=\frac{-iN_V}{4\omega_1\omega_2}\phi_{V}(\hat q)[
-M(\omega_1\omega_2+m_1m_2){\not}\epsilon+M{\not}\hat q{\not}\epsilon{\not}\hat q
-i(\omega_1m_2+m_1\omega_2){\not}P{\not}\epsilon
+iM(m_1{\not}\hat q{\not}\epsilon+m_2{\not}\epsilon{\not}\hat q)
-(\omega_1{\not}\hat q{\not}P{\not}\epsilon-\omega_2{\not}\epsilon{\not}P{\not}\hat q)]\\
\overline{\psi}^{--}_V(\hat q)=\frac{-iN_V}{4\omega_1\omega_2}\phi_{V}(\hat q)[
-M(\omega_1\omega_2+m_1m_2){\not}\epsilon+M{\not}\hat q{\not}\epsilon{\not}\hat q
+i(\omega_1m_2+m_1\omega_2){\not}P{\not}\epsilon
+iM(m_1{\not}\hat q{\not}\epsilon+m_2{\not}\epsilon{\not}\hat q)
+(\omega_1{\not}\hat q{\not}P{\not}\epsilon-\omega_2{\not}\epsilon{\not}P{\not}\hat q)]
\end{align}

The ${\not}P\overline{\psi}^{++}_{P}(\hat q'){\not}\epsilon'\Psi^{++}_{V}(\hat q)$,  ${\not}P\overline{\psi}^{++}_{P}(\hat q'){\not}\epsilon'\Psi^{--}_{V}(\hat q)$, ${\not}P\overline{\psi}^{--}_{P}(\hat q'){\not}\epsilon'\Psi^{++}_{V}(\hat q)$, and ${\not}P\overline{\psi}^{--}_{P}(\hat q'){\not}\epsilon'\Psi^{--}_{V}(\hat q)$ in the calculation ofntransition amplitude, $M_{fi}$ for $V\rightarrow P\gamma$ is done by using Eqs.(\ref{d71}) and (\ref{d72}) as:

\begin{multline}\label{vp1}
 {\not}P\overline{\psi}^{++}_{P}(\hat q'){\not}\epsilon'\Psi^{++}_{V}(\hat q)= \frac{-iN_PN_V}{16\omega_1\omega_2\omega'_1\omega'_2}\frac{M'}{M} \phi_{P}(\hat q')\phi_{V}(\hat q) \bigg[-iM(\omega'_1\omega'_2+m_1m_2+\hat q'^2)(\omega_1m_2+m_1\omega_2) {\not}P{\not}\epsilon'{\not}\epsilon{\not}P\gamma_5\\
  + iM^2 (\omega'_1\omega'_2+m_1m_2+\hat q'^2)(m_1{\not}P{\not}\epsilon'{\not}\epsilon{\not}\hat q\gamma_5+m_2{\not}P{\not}\epsilon'{\not}\hat q{\not}\epsilon\gamma_5) \\
  -iM(\omega'_1m_2+m_1\omega'_2) (\omega_1\omega_2+m_1m_2){\not}P{\not}P{\not}\epsilon'{\not}\epsilon\gamma_5\\
  +iM^3(\omega'_1m_2+m_1\omega'_2) {\not}\epsilon'{\not}\hat q{\not}\epsilon{\not}\hat q\gamma_5\\
   - iM^2 (\omega'_1m_2+m_1\omega'_2)(\omega_1{\not}\epsilon'{\not}\epsilon {\not}P{\not}\hat q\gamma_5-\omega_2{\not}\epsilon' {\not}\hat q{\not}P{\not}\epsilon\gamma_5 ) \\
 - iM^2(m_1-m_2) (\omega_1\omega_2+m_1m_2){\not}P{\not}\hat q'{\not}\epsilon'{\not}\epsilon\gamma_5\\
 +iM^2(m_1-m_2) {\not}P{\not}\hat q'{\not}\epsilon'{\not}\hat q{\not}\epsilon{\not}\hat q\gamma_5\\
 -iM(m_1-m_2)(\omega_1{\not}P{\not}\hat q'{\not}\epsilon'{\not}\epsilon {\not}P{\not}\hat q\gamma_5-\omega_2{\not}P{\not}\hat q'{\not}\epsilon'{\not}\hat q {\not}P{\not}\epsilon\gamma_5)\\
 -iM^2(\omega_1m_2+m_1\omega_2)(\omega'_1 +\omega'_2 ){\not}\hat q'{\not}\epsilon'{\not}\epsilon {\not}P\gamma_5\\
  +iM^3(m_1(\omega'_1+\omega'_2) {\not}\hat q'{\not}\epsilon'{\not}\epsilon {\not}\hat q \gamma_5
 +m_2(\omega'_1+\omega'_2) {\not}\hat q'{\not}\epsilon'{\not}\hat q{\not}\epsilon \gamma_5  \bigg],
\end{multline}

\begin{multline}\label{vp2}
 {\not}P\overline{\psi}^{--}_{P}(\hat q'){\not}\epsilon'\Psi^{--}_{V}(\hat q)=
 \frac{-iN_PN_V}{16\omega_1\omega_2\omega'_1\omega'_2}\frac{M'}{M} \phi_{P}(\hat q')\phi_{V}(\hat q) \bigg[iM(\omega'_1\omega'_2+m_1m_2+\hat q'^2)(\omega_1m_2+m_1\omega_2) {\not}P{\not}\epsilon'{\not}\epsilon{\not}P\gamma_5\\
  + iM^2 (\omega'_1\omega'_2+m_1m_2+\hat q'^2)(m_1{\not}P{\not}\epsilon'{\not}\epsilon{\not}\hat q\gamma_5+m_2{\not}P{\not}\epsilon'{\not}\hat q{\not}\epsilon\gamma_5) \\
  +iM(\omega'_1m_2+m_1\omega'_2) (\omega_1\omega_2+m_1m_2){\not}P{\not}P{\not}\epsilon'{\not}\epsilon\gamma_5\\
  -iM^3(\omega'_1m_2+m_1\omega'_2) {\not}\epsilon'{\not}\hat q{\not}\epsilon{\not}\hat q\gamma_5\\
   - iM^2 (\omega'_1m_2+m_1\omega'_2)(\omega_1{\not}\epsilon'{\not}\epsilon {\not}P{\not}\hat q\gamma_5-\omega_2{\not}\epsilon' {\not}\hat q{\not}P{\not}\epsilon\gamma_5 ) \\
 - iM^2(m_1-m_2) (\omega_1\omega_2+m_1m_2){\not}P{\not}\hat q'{\not}\epsilon'{\not}\epsilon\gamma_5\\
 +iM^2(m_1-m_2) {\not}P{\not}\hat q'{\not}\epsilon'{\not}\hat q{\not}\epsilon{\not}\hat q\gamma_5\\
 +iM(m_1-m_2)(\omega_1{\not}P{\not}\hat q'{\not}\epsilon'{\not}\epsilon {\not}P{\not}\hat q\gamma_5-\omega_2{\not}P{\not}\hat q'{\not}\epsilon'{\not}\hat q {\not}P{\not}\epsilon\gamma_5)\\
 -iM^2(\omega_1m_2+m_1\omega_2)(\omega'_1 +\omega'_2 ){\not}\hat q'{\not}\epsilon'{\not}\epsilon {\not}P\gamma_5\\
  +iM^3(m_1(\omega'_1+\omega'_2) {\not}\hat q'{\not}\epsilon'{\not}\epsilon {\not}\hat q \gamma_5
 +m_2(\omega'_1+\omega'_2) {\not}\hat q'{\not}\epsilon'{\not}\hat q{\not}\epsilon \gamma_5  \bigg],
\end{multline}

\begin{multline}\label{vp3}
 {\not}P\overline{\psi}^{++}_{P}(\hat q'){\not}\epsilon'\Psi^{--}_{V}(\hat q)= \frac{-iN_PN_V}{16\omega_1\omega_2\omega'_1\omega'_2}\frac{M'}{M} \phi_{P}(\hat q')\phi_{V}(\hat q) \bigg[-iM(\omega'_1\omega'_2+m_1m_2+\hat q'^2)(\omega_1m_2+m_1\omega_2) {\not}P{\not}\epsilon'{\not}\epsilon{\not}P\gamma_5\\
  -iM^2 (\omega'_1\omega'_2+m_1m_2+\hat q'^2)(m_1{\not}P{\not}\epsilon'{\not}\epsilon{\not}\hat q\gamma_5+m_2{\not}P{\not}\epsilon'{\not}\hat q{\not}\epsilon\gamma_5) \\
  +iM(\omega'_1m_2+m_1\omega'_2) (\omega_1\omega_2+m_1m_2){\not}P{\not}P{\not}\epsilon'{\not}\epsilon\gamma_5\\
  +iM^3(\omega'_1m_2+m_1\omega'_2) {\not}\epsilon'{\not}\hat q{\not}\epsilon{\not}\hat q\gamma_5\\
   + iM^2 (\omega'_1m_2+m_1\omega'_2)(\omega_1{\not}\epsilon'{\not}\epsilon {\not}P{\not}\hat q\gamma_5-\omega_2{\not}\epsilon' {\not}\hat q{\not}P{\not}\epsilon\gamma_5 ) \\
 - iM^2(m_1-m_2) (\omega_1\omega_2+m_1m_2){\not}P{\not}\hat q'{\not}\epsilon'{\not}\epsilon\gamma_5\\
 +iM^2(m_1-m_2) {\not}P{\not}\hat q'{\not}\epsilon'{\not}\hat q{\not}\epsilon{\not}\hat q\gamma_5\\
 +iM(m_1-m_2)(\omega_1{\not}P{\not}\hat q'{\not}\epsilon'{\not}\epsilon {\not}P{\not}\hat q\gamma_5-\omega_2{\not}P{\not}\hat q'{\not}\epsilon'{\not}\hat q {\not}P{\not}\epsilon\gamma_5)\\
 +iM^2(\omega_1m_2+m_1\omega_2)(\omega'_1 +\omega'_2 ){\not}\hat q'{\not}\epsilon'{\not}\epsilon {\not}P\gamma_5\\
  +iM^3(m_1(\omega'_1+\omega'_2) {\not}\hat q'{\not}\epsilon'{\not}\epsilon {\not}\hat q \gamma_5
 +m_2(\omega'_1+\omega'_2) {\not}\hat q'{\not}\epsilon'{\not}\hat q{\not}\epsilon \gamma_5  \bigg],
\end{multline}

and

\begin{multline} \label{vp4}
 {\not}P\overline{\psi}^{--}_{P}(\hat q'){\not}\epsilon'\Psi^{++}_{V}(\hat q)= \frac{-iN_PN_V}{16\omega_1\omega_2\omega'_1\omega'_2}\frac{M'}{M} \phi_{P}(\hat q')\phi_{V}(\hat q) \bigg[-iM(\omega'_1\omega'_2+m_1m_2+\hat q'^2)(\omega_1m_2+m_1\omega_2) {\not}P{\not}\epsilon'{\not}\epsilon{\not}P\gamma_5\\
  +iM^2 (\omega'_1\omega'_2+m_1m_2+\hat q'^2)(m_1{\not}P{\not}\epsilon'{\not}\epsilon{\not}\hat q\gamma_5+m_2{\not}P{\not}\epsilon'{\not}\hat q{\not}\epsilon\gamma_5) \\
  -iM(\omega'_1m_2+m_1\omega'_2) (\omega_1\omega_2+m_1m_2){\not}P{\not}P{\not}\epsilon'{\not}\epsilon\gamma_5\\
  -iM^3(\omega'_1m_2+m_1\omega'_2) {\not}\epsilon'{\not}\hat q{\not}\epsilon{\not}\hat q\gamma_5\\
   + iM^2 (\omega'_1m_2+m_1\omega'_2)(\omega_1{\not}\epsilon'{\not}\epsilon {\not}P{\not}\hat q\gamma_5-\omega_2{\not}\epsilon' {\not}\hat q{\not}P{\not}\epsilon\gamma_5 ) \\
 - iM^2(m_1-m_2) (\omega_1\omega_2+m_1m_2){\not}P{\not}\hat q'{\not}\epsilon'{\not}\epsilon\gamma_5\\
 +iM^2(m_1-m_2) {\not}P{\not}\hat q'{\not}\epsilon'{\not}\hat q{\not}\epsilon{\not}\hat q\gamma_5\\
 -iM(m_1-m_2)(\omega_1{\not}P{\not}\hat q'{\not}\epsilon'{\not}\epsilon {\not}P{\not}\hat q\gamma_5-\omega_2{\not}P{\not}\hat q'{\not}\epsilon'{\not}\hat q {\not}P{\not}\epsilon\gamma_5)\\
 +iM^2(\omega_1m_2+m_1\omega_2)(\omega'_1 +\omega'_2 ){\not}\hat q'{\not}\epsilon'{\not}\epsilon {\not}P\gamma_5\\
 +iM^3(m_1(\omega'_1+\omega'_2) {\not}\hat q'{\not}\epsilon'{\not}\epsilon {\not}\hat q \gamma_5
 +m_2(\omega'_1+\omega'_2) {\not}\hat q'{\not}\epsilon'{\not}\hat q{\not}\epsilon \gamma_5
 \bigg].
\end{multline}

\subsection{Radiative decays through $V\rightarrow S \gamma$} \label{A2}
The $++$ and $--$ components of scalar meson wave function in Eq.(\ref{sc1}) can be obtained through Eq.(\ref{c1}) as
\begin{align}\label{e78}
\nonumber \psi^{++}_{S}(\hat q')=\frac{-N_{S}}{4\omega'_1\omega'_2}  \phi_{S}(\hat q')[
-M(\omega'_1\omega'_2-m_1m_2+\hat q'^2)
-i(\omega'_1m_2-m_1\omega'_2){\not}P
-(\omega'_1{\not}P{\not}\hat q'-\omega'_2{\not}\hat q'{\not}P)
-iM(m_1+m_2){\not}\hat q'\\
\psi^{--}_{S}(\hat q')=\frac{-N_{S}}{4\omega'_1\omega'_2} \phi_{S}(\hat q')[
-M(\omega'_1\omega'_2-m_1m_2+\hat q'^2)
+i(\omega'_1m_2-m_1\omega'_2){\not}P
+(\omega'_1{\not}P{\not}\hat q'-\omega'_2{\not}\hat q'{\not}P)
-iM(m_1+m_2){\not}\hat q'
\end{align}

The corresponding adjoint wave functions are obtained by evaluating
$\overline{\psi}^{\pm\pm}_S(\hat q')=\gamma_4(\psi^{\pm\pm}_S(\hat q'))^+\gamma_4$ as
\begin{align}\label{e88}
\nonumber \overline{\psi}^{++}_{S}(\hat q')=\frac{-N_{S}}{4\omega'_1\omega'_2}  \phi_{S}(\hat q')[
-M(\omega'_1\omega'_2-m_1m_2+\hat q'^2)
-i(\omega'_1m_2-m_1\omega'_2){\not}P
-(\omega'_1{\not}\hat q'{\not}P-\omega'_2{\not}P{\not}\hat q')
-iM(m_1+m_2){\not}\hat q'\\
\overline{\psi}^{--}_{S}(\hat q')=\frac{-N_{S}}{4\omega'_1\omega'_2} \phi_{S}(\hat q')[
-M(\omega'_1\omega'_2-m_1m_2+\hat q'^2)
+i(\omega'_1m_2-m_1\omega'_2){\not}P
+(\omega'_1{\not}\hat q'{\not}P-\omega'_2{\not}P{\not}\hat q')
-iM(m_1+m_2){\not}\hat q'
\end{align}

The individual terms, ${\not}P\overline{\psi}^{++}_{S}(\hat q'){\not}\epsilon'\Psi^{++}_{V}(\hat q)$,  ${\not}P\overline{\psi}^{++}_{S}(\hat q'){\not}\epsilon'\Psi^{--}_{V}(\hat q)$, ${\not}P\overline{\psi}^{--}_{S}(\hat q'){\not}\epsilon'\Psi^{++}_{V}(\hat q)$, and ${\not}P\overline{\psi}^{--}_{S}(\hat q'){\not}\epsilon'\Psi^{--}_{V}(\hat q)$ in the transition amplitude, $M_{fi}$ in Eq.(\ref{7f}) can be obtained as follows:

\begin{multline}
 {\not}P\overline{\psi}^{++}_{S}(\hat q'){\not}\epsilon^{\lambda'}\psi^{++}_{V}(\hat q)
 =  \frac{-iN_{S}N_{V}}{16\omega_1\omega_2\omega'_1\omega'_2}  \phi_{S}(\hat q')\phi_{V}(\hat q) \bigg[ iM^3(\omega'_1\omega'_2-m_1m_2+\hat q'^2)(\omega_1m_2+m_1\omega_2){\not}\epsilon^{\lambda'}{\not}\epsilon^{\lambda}\\
 +iM^2 (\omega'_1\omega'_2-m_1m_2+\hat q'^2)(m_1{\not}P{\not}\epsilon^{\lambda'}{\not}\epsilon^{\lambda} {\not}\hat q+m_2{\not}P{\not}\epsilon^{\lambda'}{\not}\hat q{\not}\epsilon^{\lambda} ) \\
 +iM^3 (\omega'_1m_2-m_1\omega'_2)(\omega_1\omega_2+m_1m_2) {\not}\epsilon^{\lambda'}{\not}\epsilon^{\lambda}\\
 -iM^3(\omega'_1m_2-m_1\omega'_2){\not}\epsilon^{\lambda'}{\not}{\not}\hat q\epsilon^{\lambda}{\not}\hat q\\
 +iM^2 (\omega'_1m_2-m_1\omega'_2)(\omega_1{\not}\epsilon^{\lambda'}{\not}\epsilon^{\lambda}{\not}P {\not}\hat q -\omega_2{\not}\epsilon^{\lambda'}{\not}\hat q {\not}P{\not}\epsilon^{\lambda} ) \\
 -iM^2(\omega_1m_2+m_1\omega_2) (\omega'_1{\not}\hat q'{\not}\epsilon^{\lambda'}{\not}\epsilon^{\lambda}{\not}P+\omega'_2 {\not}\hat q'{\not}\epsilon^{\lambda'}{\not}\epsilon^{\lambda}{\not}P )\\
 +iM^3(m_1(\omega'_1+\omega'_2){\not}\hat q'{\not}\epsilon^{\lambda'}{\not}\epsilon^{\lambda}{\not}\hat q+m_2(\omega'_1+\omega'_2){\not}\hat q'{\not}\epsilon^{\lambda'}{\not}\hat q{\not}\epsilon^{\lambda})\\
 -iM^2(m_1+m_2)(\omega_1\omega_2+m_1m_2) {\not}P{\not}\hat q'{\not}\epsilon^{\lambda'} {\not}\epsilon^{\lambda}\\
 +iM^2(m_1+m_2) {\not}P{\not}\hat q'{\not}\epsilon^{\lambda'} {\not}\hat q{\not}\epsilon^{\lambda}{\not}\hat q\\
 -iM^3(m_1+m_2)(\omega_1{\not}\hat q'{\not}\epsilon^{\lambda'}{\not}\epsilon^{\lambda} {\not}\hat q-\omega_2{\not}\hat q'{\not}\epsilon^{\lambda'}{\not}\hat q{\not}\epsilon^{\lambda} ),
 \bigg]
\end{multline}

\begin{multline}
 {\not}P\overline{\psi}^{--}_{S}(\hat q'){\not}\epsilon^{\lambda'}\psi^{--}_{V}(\hat q)
 =  \frac{-iN_{S}N_{V}}{16\omega_1\omega_2\omega'_1\omega'_2}  \phi_{S}(\hat q')\phi_{V}(\hat q) \bigg[ -iM^3(\omega'_1\omega'_2-m_1m_2+\hat q'^2)(\omega_1m_2+m_1\omega_2){\not}\epsilon^{\lambda'}{\not}\epsilon^{\lambda}\\
 +iM^2 (\omega'_1\omega'_2-m_1m_2+\hat q'^2)(m_1{\not}P{\not}\epsilon^{\lambda'}{\not}\epsilon^{\lambda} {\not}\hat q+m_2{\not}P{\not}\epsilon^{\lambda'}{\not}\hat q{\not}\epsilon^{\lambda} ) \\
 -iM^3 (\omega'_1m_2-m_1\omega'_2)(\omega_1\omega_2+m_1m_2) {\not}\epsilon^{\lambda'}{\not}\epsilon^{\lambda}\\
 +iM^3(\omega'_1m_2-m_1\omega'_2){\not}\epsilon^{\lambda'}{\not}{\not}\hat q\epsilon^{\lambda}{\not}\hat q\\
 +iM^2 (\omega'_1m_2-m_1\omega'_2)(\omega_1{\not}\epsilon^{\lambda'}{\not}\epsilon^{\lambda}{\not}P {\not}\hat q -\omega_2{\not}\epsilon^{\lambda'}{\not}\hat q {\not}P{\not}\epsilon^{\lambda} ) \\
 -iM^2(\omega_1m_2+m_1\omega_2) (\omega'_1{\not}\hat q'{\not}\epsilon^{\lambda'}{\not}\epsilon^{\lambda}{\not}P+\omega'_2 {\not}\hat q'{\not}\epsilon^{\lambda'}{\not}\epsilon^{\lambda}{\not}P )\\
 -iM^3(m_1(\omega'_1+\omega'_2){\not}\hat q'{\not}\epsilon^{\lambda'}{\not}\epsilon^{\lambda}{\not}\hat q+m_2(\omega'_1+\omega'_2){\not}\hat q'{\not}\epsilon^{\lambda'}{\not}\hat q{\not}\epsilon^{\lambda})\\
 -iM^2(m_1+m_2)(\omega_1\omega_2+m_1m_2) {\not}P{\not}\hat q'{\not}\epsilon^{\lambda'} {\not}\epsilon^{\lambda}\\
 +iM^2(m_1+m_2) {\not}P{\not}\hat q'{\not}\epsilon^{\lambda'} {\not}\hat q{\not}\epsilon^{\lambda}{\not}\hat q\\
 +iM^3(m_1+m_2)(\omega_1{\not}\hat q'{\not}\epsilon^{\lambda'}{\not}\epsilon^{\lambda} {\not}\hat q-\omega_2{\not}\hat q'{\not}\epsilon^{\lambda'}{\not}\hat q{\not}\epsilon^{\lambda} ),
 \bigg]
\end{multline}

\begin{multline}
 {\not}P\overline{\psi}^{++}_{S}(\hat q'){\not}\epsilon^{\lambda'}\psi^{--}_{V}(\hat q)
 =  \frac{-iN_{S}N_{V}}{16\omega_1\omega_2\omega'_1\omega'_2}  \phi_{S}(\hat q')\phi_{V}(\hat q) \bigg[ -iM^3(\omega'_1\omega'_2-m_1m_2+\hat q'^2)(\omega_1m_2+m_1\omega_2){\not}\epsilon^{\lambda'}{\not}\epsilon^{\lambda}\\
 +iM^2 (\omega'_1\omega'_2-m_1m_2+\hat q'^2)(m_1{\not}P{\not}\epsilon^{\lambda'}{\not}\epsilon^{\lambda} {\not}\hat q+m_2{\not}P{\not}\epsilon^{\lambda'}{\not}\hat q{\not}\epsilon^{\lambda} ) \\
 +iM^3 (\omega'_1m_2-m_1\omega'_2)(\omega_1\omega_2+m_1m_2) {\not}\epsilon^{\lambda'}{\not}\epsilon^{\lambda}\\
 -iM^3(\omega'_1m_2-m_1\omega'_2){\not}\epsilon^{\lambda'}{\not}{\not}\hat q\epsilon^{\lambda}{\not}\hat q\\
 -iM^2 (\omega'_1m_2-m_1\omega'_2)(\omega_1{\not}\epsilon^{\lambda'}{\not}\epsilon^{\lambda}{\not}P {\not}\hat q -\omega_2{\not}\epsilon^{\lambda'}{\not}\hat q {\not}P{\not}\epsilon^{\lambda} ) \\
 +iM^2(\omega_1m_2+m_1\omega_2) (\omega'_1{\not}\hat q'{\not}\epsilon^{\lambda'}{\not}\epsilon^{\lambda}{\not}P+\omega'_2 {\not}\hat q'{\not}\epsilon^{\lambda'}{\not}\epsilon^{\lambda}{\not}P )\\
 +iM^3(m_1(\omega'_1+\omega'_2){\not}\hat q'{\not}\epsilon^{\lambda'}{\not}\epsilon^{\lambda}{\not}\hat q+m_2(\omega'_1+\omega'_2){\not}\hat q'{\not}\epsilon^{\lambda'}{\not}\hat q{\not}\epsilon^{\lambda})\\
 -iM^2(m_1+m_2)(\omega_1\omega_2+m_1m_2) {\not}P{\not}\hat q'{\not}\epsilon^{\lambda'} {\not}\epsilon^{\lambda}\\
 +iM^2(m_1+m_2) {\not}P{\not}\hat q'{\not}\epsilon^{\lambda'} {\not}\hat q{\not}\epsilon^{\lambda}{\not}\hat q\\
 +iM^3(m_1+m_2)(\omega_1{\not}\hat q'{\not}\epsilon^{\lambda'}{\not}\epsilon^{\lambda} {\not}\hat q-\omega_2{\not}\hat q'{\not}\epsilon^{\lambda'}{\not}\hat q{\not}\epsilon^{\lambda} ),
 \bigg]
\end{multline}

 and,
\begin{multline}
 {\not}P\overline{\psi}^{--}_{S}(\hat q'){\not}\epsilon^{\lambda'}\psi^{++}_{V}(\hat q)
 =  \frac{-iN_{S}N_{V}}{16\omega_1\omega_2\omega'_1\omega'_2}  \phi_{S}(\hat q')\phi_{V}(\hat q) \bigg[ iM^3(\omega'_1\omega'_2-m_1m_2+\hat q'^2)(\omega_1m_2+m_1\omega_2){\not}\epsilon^{\lambda'}{\not}\epsilon^{\lambda}\\
 +iM^2 (\omega'_1\omega'_2-m_1m_2+\hat q'^2)(m_1{\not}P{\not}\epsilon^{\lambda'}{\not}\epsilon^{\lambda} {\not}\hat q+m_2{\not}P{\not}\epsilon^{\lambda'}{\not}\hat q{\not}\epsilon^{\lambda} ) \\
 -iM^3 (\omega'_1m_2-m_1\omega'_2)(\omega_1\omega_2+m_1m_2) {\not}\epsilon^{\lambda'}{\not}\epsilon^{\lambda}\\
 +iM^3(\omega'_1m_2-m_1\omega'_2){\not}\epsilon^{\lambda'}{\not}{\not}\hat q\epsilon^{\lambda}{\not}\hat q\\
 -iM^2 (\omega'_1m_2-m_1\omega'_2)(\omega_1{\not}\epsilon^{\lambda'}{\not}\epsilon^{\lambda}{\not}P {\not}\hat q -\omega_2{\not}\epsilon^{\lambda'}{\not}\hat q {\not}P{\not}\epsilon^{\lambda} ) \\
 +iM^2(\omega_1m_2+m_1\omega_2) (\omega'_1{\not}\hat q'{\not}\epsilon^{\lambda'}{\not}\epsilon^{\lambda}{\not}P+\omega'_2 {\not}\hat q'{\not}\epsilon^{\lambda'}{\not}\epsilon^{\lambda}{\not}P )\\
 -iM^3(m_1(\omega'_1+\omega'_2){\not}\hat q'{\not}\epsilon^{\lambda'}{\not}\epsilon^{\lambda}{\not}\hat q+m_2(\omega'_1+\omega'_2){\not}\hat q'{\not}\epsilon^{\lambda'}{\not}\hat q{\not}\epsilon^{\lambda})\\
 -iM^2(m_1+m_2)(\omega_1\omega_2+m_1m_2) {\not}P{\not}\hat q'{\not}\epsilon^{\lambda'} {\not}\epsilon^{\lambda}\\
 +iM^2(m_1+m_2) {\not}P{\not}\hat q'{\not}\epsilon^{\lambda'} {\not}\hat q{\not}\epsilon^{\lambda}{\not}\hat q\\
 -iM^3(m_1+m_2)(\omega_1{\not}\hat q'{\not}\epsilon^{\lambda'}{\not}\epsilon^{\lambda} {\not}\hat q-\omega_2{\not}\hat q'{\not}\epsilon^{\lambda'}{\not}\hat q{\not}\epsilon^{\lambda} ).
 \bigg]
\end{multline}

\end{document}